%% file: main.tex
\pdfoutput=1
 
\RequirePackage{fix-cm}
\documentclass[twocolumn]{svjour3}          

\makeatletter
\if@twocolumn
  \renewcommand\normalsize{%
   \@setfontsize\normalsize\@xpt{12.5pt}%
   \abovedisplayskip=3 mm plus6pt minus 4pt
   \belowdisplayskip=3 mm plus6pt minus 4pt
   \abovedisplayshortskip=0.0 mm plus6pt
   \belowdisplayshortskip=2 mm plus4pt minus 4pt
   \let\@listi\@listI}%

  \renewcommand\small{%
   \@setfontsize\small{8.5pt}\@xpt
   \abovedisplayskip 8.5\p@ \@plus3\p@ \@minus4\p@
   \abovedisplayshortskip \z@ \@plus2\p@
   \belowdisplayshortskip 4\p@ \@plus2\p@ \@minus2\p@
   \def\@listi{\leftmargin\leftmargini
               \parsep 0\p@ \@plus1\p@ \@minus\p@
               \topsep 4\p@ \@plus2\p@ \@minus4\p@
               \itemsep0\p@}%
   \belowdisplayskip \abovedisplayskip}

\else
  \if@smallext
   \renewcommand\normalsize{%
   \@setfontsize\normalsize\@xpt\@xiipt
   \abovedisplayskip=3 mm plus6pt minus 4pt
   \belowdisplayskip=3 mm plus6pt minus 4pt
   \abovedisplayshortskip=0.0 mm plus6pt
   \belowdisplayshortskip=2 mm plus4pt minus 4pt
   \let\@listi\@listI}%

  \renewcommand\small{%
   \@setfontsize\small\@viiipt{9.5pt}%
   \abovedisplayskip 8.5\p@ \@plus3\p@ \@minus4\p@
   \abovedisplayshortskip \z@ \@plus2\p@
   \belowdisplayshortskip 4\p@ \@plus2\p@ \@minus2\p@
   \def\@listi{\leftmargin\leftmargini
               \parsep 0\p@ \@plus1\p@ \@minus\p@
               \topsep 4\p@ \@plus2\p@ \@minus4\p@
               \itemsep0\p@}%
   \belowdisplayskip \abovedisplayskip}
 \else
  \renewcommand\normalsize{%
   \@setfontsize\normalsize{9.5pt}{11.5pt}%
   \abovedisplayskip=3 mm plus6pt minus 4pt
   \belowdisplayskip=3 mm plus6pt minus 4pt
   \abovedisplayshortskip=0.0 mm plus6pt
   \belowdisplayshortskip=2 mm plus4pt minus 4pt
   \let\@listi\@listI}%

  \renewcommand\small{%
   \@setfontsize\small\@viiipt{9.25pt}%
   \abovedisplayskip 8.5\p@ \@plus3\p@ \@minus4\p@
   \abovedisplayshortskip \z@ \@plus2\p@
   \belowdisplayshortskip 4\p@ \@plus2\p@ \@minus2\p@
   \def\@listi{\leftmargin\leftmargini
               \parsep 0\p@ \@plus1\p@ \@minus\p@
               \topsep 4\p@ \@plus2\p@ \@minus4\p@
               \itemsep0\p@}%
   \belowdisplayskip \abovedisplayskip}
  \fi
\fi

\makeatother

\smartqed  
\usepackage{graphicx}
\usepackage{academicons}
\usepackage{xcolor}
\usepackage{svg}
\usepackage{url}
\usepackage{listings}
\usepackage{microtype}

\newcommand{\orcid}[1]{\href{https://orcid.org/#1}{\includegraphics[width=10pt]{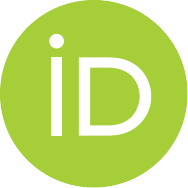}}}

\usepackage{hyperref}

\begin{document}

\title{A GPU-based Kalman Filter for Track Fitting
}


\author{Xiaocong Ai~\orcid{0000-0003-3856-2415} \and
      Georgiana Mania~\orcid{0000-0001-7536-5336} \and
      Heather M. Gray~\orcid{0000-0002-5293-4716} \and
      Michael~Kuhn~\orcid{0000-0001-8167-8574} \and
      Nicholas Styles~\orcid{0000-0001-6976-9457}
}


\institute{
X. Ai \at
Deutsches Elektronen-Synchrotron, Hamburg, Germany\\
\email{xiaocong.ai@desy.de}
\and 
G. Mania \at
Deutsches Elektronen-Synchrotron, Hamburg, Germany and University of Hamburg, Informatics Department, Hamburg, Germany
\and
H.M. Gray \at
University of California, Berkeley CA, USA  and Physics Division, Lawrence Berkeley National Laboratory
\and
M. Kuhn \at
Otto von Guericke University Magdeburg,
Faculty of Computer Science,
Magdeburg,
Germany
\and 
N. Styles \at
Deutsches Elektronen-Synchrotron, Hamburg, Germany
}

\date{Received: date / Accepted: date}

\maketitle
\begin{abstract}
Computing centres, including those used to process High-Energy Physics data and simulations, are increasingly providing significant fractions of their computing resources through hardware architectures other than x86 CPUs, with GPUs being a common alternative. GPUs can provide excellent computational performance at a good price point for tasks that can be suitably parallelized. Charged particle (track) reconstruction is a computationally expensive component of HEP data reconstruction, and thus needs to use available resources in an efficient way. In this paper, an implementation of Kalman filter-based track fitting using CUDA and running on GPUs is presented. This utilizes the ACTS (A Common Tracking Software) toolkit; an open source and experiment-independent toolkit for track reconstruction. The implementation details and parallelization approach are described, along with the specific challenges for such an implementation. Detailed performance benchmarking results are discussed, which show encouraging performance gains over a CPU-based implementation for representative configurations. Finally, a perspective on the challenges and future directions for these studies is outlined. These include more complex and realistic scenarios which can be studied, and anticipated developments to software frameworks and standards which may open up possibilities for greater flexibility and improved performance.

\end{abstract}

\input{introduction}

\input{software_FW}

\input{parallelizationTech}

\input{GPU_impl}

\input{performance}

\input{discussion}

\input{conclusion}

\begin{acknowledgements}
We would like to thank Dr. Attila Krasznahorkay (CERN), Dr. Charles Leggett (Lawrence Berkeley National Laboratory) and Dr. Andreas Salzburger (CERN) for their careful reading of the manuscript and their helpful comments and suggestions.
This work used the grid computational resources operated at Deutsches Elektronen-Synchrotron (DESY), Hamburg, Germany and at the National Energy Research Scientific Computing Center (NERSC), a U.S. Department of Energy Office of Science User Facility located at Lawrence Berkeley National Laboratory, operated under Contract No. DE-AC02-05CH11231.
We acknowledge the support  by the National Science Foundation (NSF) and Data Science in Hamburg - HELMHOLTZ Graduate School for the Structure of Matter (DASHH).

\end{acknowledgements}

\section*{Declarations}
\paragraph{Funding}
This work was funded by the NSF under Cooperative Agreement OAC-1836650, and supported by DASHH with the Grant-No. HIDSS-0002. 

\paragraph{Conflict of interest}
The authors declare that they have no conflict of interest.

\paragraph{Availability of data and material} 
Not applicable. No associated data except for code.

\paragraph{Code availability}
The code used for this research (including a Singularity container for reproducibility) is available open source~\cite{xiaocong_ai_2021_4693389}.

%
%

\bibliographystyle{spmpsci}      
\bibliography{references}   

%
%

\end{document}

%% file: introduction.tex
\section{Introduction}
\label{sec:intro}

The reconstruction of the trajectories of charged particles for High-Energy Physics (HEP) experiments is a very computationally demanding task, which is performed both when selecting events in real time with the online \emph{trigger} and during the subsequent high-precision offline reconstruction of events for physics analysis. The most commonly used techniques are adaptive methods based on the Kalman filter~\cite{Billoir:1983mz,Fruhwirth:1987fm}, which account for the trajectories of charged particles in magnetic fields and the energy loss of charged particles in the detector material. See Ref.~\cite{RevModPhys.82.1419} for a review. As the execution time of such algorithms explodes combinatorially with the number of charged particles, the advent of the upgrade to the Large Hadron Collider (LHC), the High-Luminosity LHC (HL-LHC), portends an even greater challenge, with events containing up to 10,000 tracks.

For many years, HEP has been relying on Moore's Law~\cite{moore}, the observation that the number of transistors on an integrated circuit doubles approximately every two years. As the circuits have begun to approach intrinsic limits in terms of density and power, Moore's Law has begun to slow, further complicating potential performance improvements~\cite{Shalf2020TheFO}. In addition, other computing architectures have become increasingly powerful and hence popular, such as graphical processing units (GPUs) and field programmable gate arrays (FPGAs). Therefore there has been a shift towards achieving speed improvements by adding additional cores, particularly at high-performance computing centers. These many-core systems require highly parallel code to be fully exploited, requiring additional knowledge from software developers. Moreover, much of the existing code for high-energy physics experiments is not well-suited to such architectures and hence requires significant development and adaptation to be able to exploit them. 

Porting algorithms to GPUs typically requires specialized code redesign and optimization, but performance gains through vectorization using Single Instruction Multiple Data (SIMD) instructions, and parallelization using many-core CPU architectures often require less significant changes to the code base. Several HEP experiments have leveraged the power of many-core systems for real-time online and/or offline track reconstruction~\cite{Cerati_2014,Cerati_2017,Cerati_2020,Lantz_2020,Kisel_2018}. These studies have demonstrated good scalability of the throughput of events per second with the number of CPU cores. GPU-accelerated track reconstruction has also been studied. For example, both the ALICE~\cite{ALICE_2008} and LHCb~\cite{LHCb_2008} experiments at the LHC have proposed a GPU-based High Level Trigger (HLT) to handle the much increased data rate expected during Run 3 of LHC~\cite{rohr2018track,Rohr_2019,Aaij2020}. In particular, LHCb has implemented a fully GPU-based high-throughput HLT framework, which processes a data rate of up to 40 Tbit/s using approximately 500 GPUs~\cite{Aaij2020}. In these studies, ALICE and LHCb used a simplified or parametrized Kalman filter for track fitting for maximum speed with some impact on track resolution compared to offline track reconstruction using a full Kalman filter. The level of resolution loss is either acceptable for the online identification of interesting events for further offline analysis~\cite{Aaij2020} or is recovered through dedicated optimization of the HLT tracking algorithms~\cite{rohr2018track}. Initial studies of porting a full Kalman filter to GPUs can be found in Ref.~\cite{Cerati_2017}. Other track finding algorithms such as the Cellular Automaton and Hough Transforms have also been studied on GPUs~\cite{Funke_2014,rinaldi2015gpgpu}. GPUs are also used for accelerating other steps of online event processing at HEP experiments, e.g. cluster finding~\cite{Aaij2020,bocci2020heterogeneous}, vertex reconstruction~\cite{Bruch_2017} and event selection~\cite{Sen_2015} and Ref.~\cite{Bruch_2020} presents a recent review of applications of GPUs for online event processing in HEP. The trend is generally towards bringing the full reconstruction chain to GPUs in order to minimize the penalties from intermediate data transfer between host and GPUs (see Ref.~\cite{Aaij2020,bocci2020heterogeneous}). Beyond HEP, GPU-accelerated Kalman filtering has been explored for a range of applications~\cite{Huang_2011,Xu_2016}. However, these use cases tend to focus on much larger (up to three orders of magnitude) matrix sizes than are typical in HEP applications, and so the direct applicability is limited.

We present a proof-of-concept of a full Kalman filter algorithm on GPUs utilizing A Common Tracking Software (ACTS)~\cite{Gumpert:2243297,Ai:2019kze,Gessinger:2020nne,Ai:2020jbw,ai2021common}, which provides a toolkit of algorithms for track reconstruction within a generic, framework- and experiment-independent software package. Detailed studies of the physics and technical performance are presented for two different GPU architectures and compared to performance on CPUs. In particular, we identify and discuss the key challenges in the implementation and highlight future directions towards the development of an even more performant full Kalman filter algorithm.

%% file: software_FW.tex
\section{The Kalman Filter and ACTS}
\label{sec:kf}

Track reconstruction is typically a multi-stage procedure, wherein candidates can be rejected at each stage. This approach allows high reconstruction efficiency and purity to be achieved in the final output collection, while reducing the overhead from processing unwanted candidates further than necessary. It starts from measurements (deposited energy in sensitive elements of the detector) and combines them in various configurations (including appropriate calibrations at various stages) to form plausible candidate trajectories. Accurate estimations of the parameters which define the mathematical form used to describe the trajectory are then made. 

After any required pre-processing of the raw measurements, a typical first step is \emph{Seeding}, in which small sets of compatible measurements are grouped using simple criteria and an initial trajectory estimate made. Seeds passing requirements can then be used as the basis for \emph{Track Finding}, in which additional compatible measurements are added to the trajectory through the detector. Once the full set of measurements for the trajectory is obtained, a \emph{Track Fitting} step can be performed, in order to precisely estimate the parameters and their covariance.

A commonly used approach and important tool in many track reconstruction applications is the Kalman filter~\cite{Billoir:1983mz,Fruhwirth:1987fm}. Developed in the late 1950s, the initial application of the Kalman filter procedure was in ballistics~\cite{kalman1960}, where it allowed telemetry data for the heading and acceleration of the projectile to be combined with information on its location. The generalized procedure, in which measurements are combined with predictions based on an underlying model, results in state estimates more precise than either measurements or predictions alone, and has since been very widely used in many fields.

\begin{figure*}[ht]
\centering
\includegraphics[width=\textwidth]{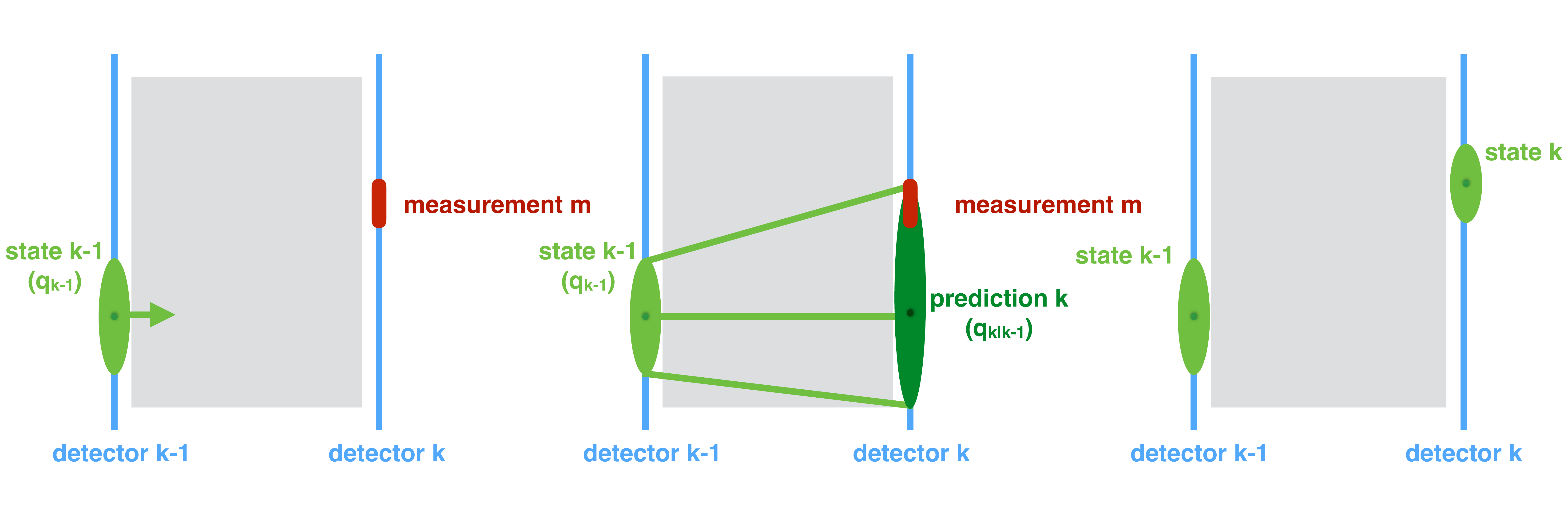}
\caption{Illustration of the different steps of the Kalman filter using a simplified detector model consisting of only two layers. (Left) The track state at the $(k-1)$th surface is indicated with the light green ellipse, and a measurement on the $k$th surface is indicated in red. (Center) A prediction is made for the track state on the $k$th surface and the size of the prediction is indicated with the dark green ellipse. (Right) The track state on the $k$th surface is updated by including the measurement on the $k$th surface \label{fig:kalmanfilter}}
\end{figure*}

Within track reconstruction, a typical Kalman filter \emph{step} would proceed as follows (see Fig.~\ref{fig:kalmanfilter}):

\begin{enumerate}
   \item An initial estimate of the track state (i.e. helix parameters) at a given position is taken as the starting point
   \item This track state is propagated according to the track model on to the next \emph{Measurement Surface} (i.e. the reference plane of a sensitive detector), providing a prediction of track state on this surface
   \item The prediction is combined with the measurement at this surface, if present, either through a weighted average or the so-called \emph{Gain Matrix} formalism forming a new track state which is used to update the initial estimate
   \item This new estimate is then used for further Kalman steps, up to the end of the trajectory
\end{enumerate}

\begin{figure}[!htb]
\centering
\includegraphics[width=\linewidth]{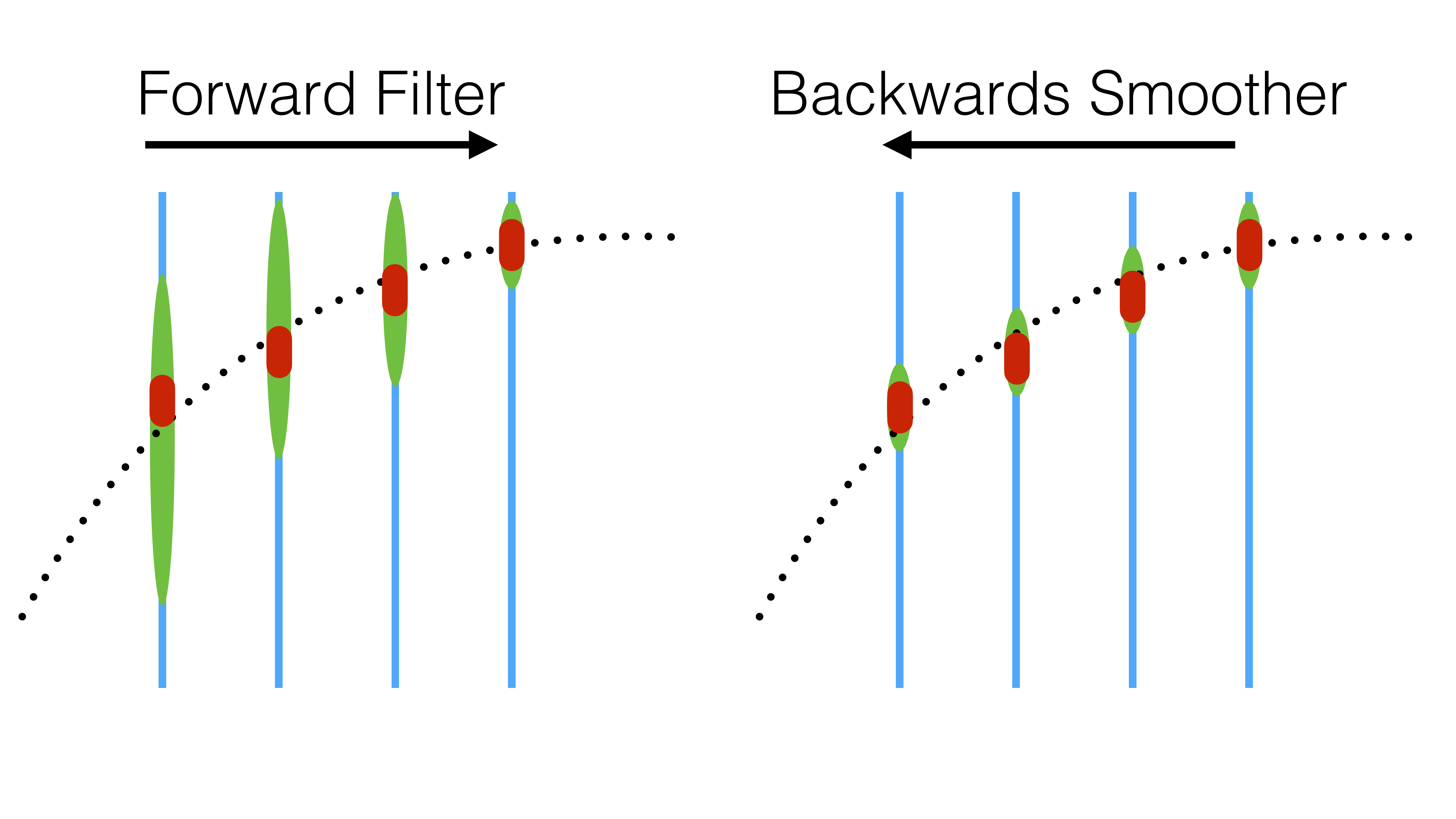}
\caption{Illustration of a forward filter and a backwards smoother on a simplified four layer detector geometry. The red points indicate the measurements and their uncertainties on each layer. The green points indicate the predictions. The predictions from the forward filter (left) are obtained when the filter is run from left to right. The predictions from the backwards filter are obtained during a second pass of the filter when it is run from right to left\label{fig:smoother}}
\end{figure}

The Kalman procedure has the property that the next state estimate can be determined from the one preceding it. While this is a useful property in many cases, as it requires no `history' to be stored, it has the consequence that only the final state contains the full information about all the steps preceding it, and therefore the best possible precision. To allow the prior track states to benefit from this information (e.g. to allow a $\chi ^2$ quality metric to be defined based on measurement residuals), an additional stage is needed. This \emph{Smoother} stage can be performed using one of two approaches: either using essentially the same procedure as the forward Kalman filter but in reverse direction as illustrated in Fig.~\ref{fig:smoother} or using the Rauch-Tung-Striebel (RTS) smoother \cite{RTS} formalism with the stored Jacobians between states calculated during the forward Kalman filter steps. The latter approach does not require a second propagation of the track parameters and is therefore expected to have better timing performance.

ACTS has its origins in the track reconstruction algorithms used by the ATLAS experiment~\cite{Aad:1129811}. In addition to a tracking toolkit, ACTS also includes a fast simulation package. The ACTS code is designed to be inherently thread-safe to support parallel code execution and the data structures are vectorised. The implementation has been designed to be fully agnostic to detection technologies, detector design, and software frameworks so that it can be used by a range of experiments. The Eigen library~\cite{Guennebaud:2010aa} is used for algebra operations. In addition, ACTS is designed to be an R\&D platform for the development of new algorithms and the porting of existing algorithms to new hardware platforms. See Ref.\cite{Ai:2020jbw,ai2021common} for further details.

While various representations of trajectories are possible, in this paper we will focus on helical trajectories of charged particles in a solenoidal magnetic field described using the following parameters:
\begin{itemize}
\item Two parameters $loc_0$ and $loc_1$ describing the spatial coordinates represented in the local frame of the measurement plane. In the special case of describing the parameters at a perigee surface\footnote{A surface defined at the point of closest approach to a reference point, for example the nominal interaction point in a particle collider}, these become the transverse and longitudinal impact parameters $d_0$ and $z_0$ respectively.
\item The polar and azimuthal angles of the particle momentum vector direction, $\phi$ and $\theta$, at that point.
\item A curvature parameter, expressed as the ratio of charge to momentum $\frac{q}{p}$.
\item The time $t$.
\end{itemize}
Both the backwards-propagation and the RTS Kalman smoothing approaches are available within ACTS. The latter approach is used for the performance studies in this paper.

%% file: parallelizationTech.tex
\section{Parallelization and Offloading Techniques}

There is a wide range of tools and frameworks available that can improve the runtime performance of scientific code via parallelization and offloading.
Two of the most widely used frameworks are Open Multi-Processing (OpenMP)~\cite{660313} and Compute Unified Device Architecture (CUDA)~\cite{cuda_doc}.
While the former traditionally allows parallelization on CPU systems via multiple threads, the latter is used to offload parts of the code to massively-parallel Nvidia GPUs.

\subsection{OpenMP}

OpenMP is a compiler-based high-level approach for thread parallelization on shared memory architectures.
One of its outstanding features is that it is very easy to use and does not require knowledge of threading and operating system internals~\cite{DBLP:journals/ieeecc/Clark98}.
It is available for C, C++ and Fortran, some of the most-widely used programming languages for scientific computing.
OpenMP achieves its simplicity by being integrated into the compiler, which facilitates the parallelization of applications.
Compiler support is widely available, which allows it to be used on personal computers as well as supercomputers.
Applications are annotated with so-called \emph{pragmas}, and turned into parallel code by the compiler; for instance, a simple \verb!#pragma omp parallel for! pragma instructs the compiler to parallelize the subsequent \emph{for} loop.
OpenMP takes care of thread management and scheduling as well as data decomposition, which allows developers to focus on the problem they want to solve.

One of OpenMP's drawbacks has been its focus on CPU-based parallelism.
However, it has recently been extended with improved offloading functionality that allows the compiler to offload certain parts of an application to accelerators such as GPUs and FPGAs.
Consequently, OpenMP now can target both CPUs and GPUs, which offers better portability than vendor-specific approaches such as CUDA~\cite{DBLP:conf/iwomp/DaleyAWW20}.

\subsection{CUDA}
\label{cuda}

CUDA is a parallel computing platform and application programming interface introduced by Nvidia for their line of GPUs~\cite{DBLP:journals/queue/NickollsBGS08}.
It allows highly-parallel GPUs to be used for general-purpose computations such as those common in high-energy physics.
It is available for a range of programming languages, including C, C++, and Fortran.
Wrappers are also available for additional programming languages, such as Python, R, Julia and many others.

GPUs are very specialized processing units and feature a high number of computing cores, which can be leveraged for scientific computations.
Programs are offloaded to the GPUs in the form of so-called \emph{compute kernels}, that is, single functions and their associated data.
A kernel executes in parallel across a set of threads, which can use per-thread registers.
Moreover, threads are aggregated into so-called warps that are executed concurrently.
Several of these warps can be grouped into a thread block, which has access to a fast region of shared memory that all threads within the block can access.
Finally, thread blocks can be combined into grids by the programmer.
Thread blocks in a grid can only share data via global memory. Details of the CUDA programming model can be found in Ref.~\cite{cuda_program}.

CUDA extends existing languages and requires dedicated compilers (\texttt{nvcc} for C/C++ and \texttt{nvfortran} for Fortran).
While it allows for the optimal use of Nvidia GPUs, it is not portable and cannot be used for GPUs produced by other vendors.
However, there have been attempts to provide abstraction layers or conversion tools for other approaches to be able to run CUDA code via OpenCL~\cite{DBLP:journals/pc/DuWLTPD12,DBLP:conf/iwocl/BabejJ20}.
There are also a variety of libraries that automatically offload compute-intensive operations to the GPUs.
Examples include libraries such as cuBLAS for linear algebra and cuFFT for fast-fourier transforms~\cite{7476520}.

Competing approaches including OpenACC~\cite{DBLP:conf/sc/HerdmanGPBMJ14} and OpenCL are available but have not been not as widely adopted so far.
For the parallel GPU implementation presented in this paper, we have chosen CUDA because it is the de-facto standard for GPU-accelerated code and is widely supported.
Our attempts to develop a GPU-accelerated solution using OpenMP were not successful so far due to offloading support in OpenMP still being in an early stage of development.

%% file: GPU_impl.tex
\section{GPU Implementation}
\label{sec:gpu_imple}
In this section, implementation details of the Kalman filter used to run track fitting on both CPUs and GPUs are presented. The code can be found in Ref.~\cite{xiaocong_ai_2021_4693389}.

\subsection{Parallelization Strategy}
As discussed in Sect.~\ref{sec:kf}, track fitting is the step in the track reconstruction chain that precisely estimates the reconstructed track parameters and the associated covariance matrices. If track fitting is performed sequentially in a single event, the execution time will increase almost linearly with increasing track multiplicity. However, the dependence of the track fitting execution time on the number of tracks weakens if the track fitting can be parallelized. The implementation of a track-level parallel strategy is straightforward, since the track fitting for each reconstructed track is completely independent. In addition, the algorithm can be parallelized \emph{within} a track fit, i.e.~intra-track parallelization. Possible gains come from the matrix operations, e.g. the transportation of the track parameters in a magnetic field and the Kalman filter update and smoothing, which are computationally expensive and have to be repeated for all the propagation steps and measurements by a total of up to $\mathcal{O}(10)$ times per track. However, in practise only very limited intra-track parallelization for those matrix operations can be achieved by using multiple threads, because the sizes of the matrices in one operation are usually relatively small. For example, the largest size matrix operated on in a single track fit in ACTS is the covariance matrix of the track parameters represented in the global coordinate system, which is of size 8$\times$8.

This paper discusses both parallelization strategies for track fitting on GPUs:

\begin{enumerate}
    \item Track-level parallelization: Track fitting for different tracks is executed in parallel using different CUDA threads (or blocks if further intra-track parallelization is used).
    \item Intra-track parallelization: The matrix operations involved in a single track fit are parallelized as much as possible using multiple threads within a single CUDA block. In this case, the block shared memory is used for the objects relevant with one track fit.
\end{enumerate}

The transportation of the track parameters and their associated covariance matrices in a magnetic field requires a numerical solution to the equation of particle motion. The adaptive Runge-Kutta-Nyström~\cite{Myrheim:1979ng} method is used to transport the track parameters in ACTS. When extrapolating the track parameters from one measurement point to the next, the covariance of the track parameters is updated with the transport Jacobian between the measurement points. Because the track parameters are represented in the local coordinate frame of the detector, the transform Jacobians between local and global track parameters at the two measurement points have to be applied. If the fitting is performed using one CUDA block per track, the matrix multiplication for the covariance transport can be parallelized using multiple threads.

\subsection{CUDA Considerations and Limitations}
Various CUDA programming requirements have consequences for the problem-specific factors that shape the parallelization strategies, and thus have an impact on the final implementation. The most challenging ones are detailed next.

\subsubsection{Polymorphism}
Virtual functions cannot be called inside a CUDA kernel unless the objects are constructed there. The Curiously Recurring Template Pattern (CRTP) is a C++ design pattern that emulates the behaviour of dynamic polymorphism through having a base class which is a template specialization for the derived class itself.

The ACTS Kalman filter is designed to be independent of the detector's tracking geometry, which could contain surfaces of different concrete types for different tracking detectors. To realize this design pattern on accelerators, the surfaces are implemented with CRTP, instantiated outside of the Kalman filter, and fed to the algorithm. CRTP is successfully used to define the surfaces as shown in the code sample in Listing~\ref{lst:poly}.

\begin{lstlisting}[language=C++, caption={Function definition in Surface base class using CRTP}, label={lst:poly}]
template <typename Derived>
inline const typename Derived::SurfaceBoundsType
    *Surface ::bounds() const {
  return static_cast<const Derived *>(this)->bounds();
}
\end{lstlisting}

\subsubsection{Thread Memory Limitations}
The amount of memory available for a thread, which includes the stack frame size and the maximum number of registers, is automatically configured by the CUDA runtime environment based on device properties including the total amount of shared memory and the cache sizes, and also on the number of parallel threads per execution block. Because the GPU's performance gain is based on the ability to run thousands of threads in parallel, this limits considerably the amount of memory available per thread compared to the CPU. This memory limitation is a major concern for recursive functions, which have to be reimplemented using an iterative approach.

\subsubsection{Limited Support for Linear Algebra Libraries}
\label{subsubsec:linear_algebra_support}
In the Kalman smoothing, the track parameter covariance matrix needs to be inverted to calculate the gain matrix. While the Eigen-based matrix class has a member function that returns the inverse matrix, this method is not supported in the GPU kernels.
Also CUDA 10.0 discontinued the support for invoking cuBLAS functions from within the device kernels through \verb!cublas_device! routine~\cite{no-cublas}.
Since at the moment neither of these two linear algebra libraries provides a solution for our scenario, a customized matrix inverter is implemented whose performance impact is discussed in Sect.~\ref{sec:perf}.

\subsubsection{Precision and Rounding}
Heterogeneous resources produce slightly different results due to different approaches to floating point arithmetic and rounding. CPUs typically promote float operands to doubles when possible, perform the operations in double precision and then truncate the result to simple precision.
Moreover, the x86 floating point units use extended double precision registers (80-bit), while CUDA limits the register sizes to 32-bit and 64-bit as described by the IEEE standard 754-2008~\cite{ieee_754,cuda_doc_fp}. 

To mitigate these effects, the customized matrix inverter is always configured to perform the algebra operations using double precision. Floating point precision for other algebra operations is used for benchmarking the performance, and the comparison between the results with float and double precision is discussed in Sect.~\ref{subsec:tech_perf}.

\subsection{Data Structure and Transfer}
\label{subsec:data_transfer}
Detector geometry information is a fundamental component for track parameter propagation and the integration of material effects during the track reconstruction. Because track reconstruction with a realistic detector description as used in full detector simulation requires significant computational resources, a simplified detector geometry, the so-called \emph{tracking geometry}, is used during the track reconstruction in ACTS for fast navigation and extrapolation of tracks. The basic geometrical component of the tracking geometry in ACTS is the surface. The surface object carries information about its geometrical orientation, shape and boundary, the material approximated from a full detector geometry, and a unique hierarchical geometry identifier.

Magnetic fields are used for measuring the momentum of charged particles at high-energy physics experiments. When a charged particle passes through a magnetic field, it bends with the degree of bending inversely proportional to its momentum. In this paper, a constant magnetic field represented as an Eigen matrix of size 3$\times$1 is used for the performance studies presented. It should be noted that this represents a significant simplification with respect to an inhomogeneous magnetic field, as found in many experiments, for which position-specific field information may need to be stored and retrieved, i.e. ~more memory might be required. The description of inhomogeneous or nonparametric magnetic fields however is also possible within ACTS.

\begin{figure}[b]
  \centering
  \includegraphics[width=1.0\linewidth]{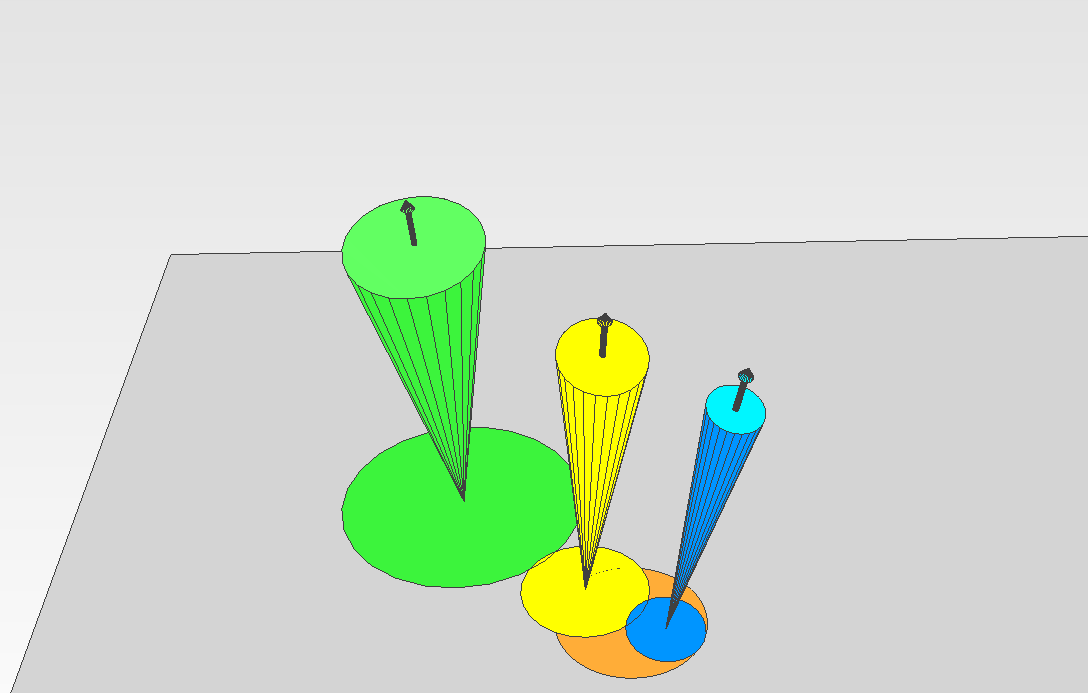}
  \caption{Demonstration of a track state with a measurement (orange), a predicted track parameter (green), a filtered track parameter (yellow) and a smoothed track parameter (blue) on a plane surface (gray). The covariance matrix of the local coordinates for both the measurement and the fitted track parameters is represented by an ellipse. The momentum direction of the fitted track parameters is represented by an arrow with its covariance matrix represented by a cone oriented in the direction of the arrow}
  \label{fig:vis_ts}
\end{figure}

The ACTS tracking Event Data Model (EDM) includes classes to describe tracking objects such as measurements and track parameters, which are represented with surfaces. The EDM for the track state is designed for the Kalman filter. It consists of a measurement, a predicted track parameter, a filtered track parameter and a smoothed track parameter, all located on a surface. Figure~\ref{fig:vis_ts} shows a track state on a surface.
During the construction of the detector geometry, a unique geometrical identifier is assigned to each detector surface. By storing the geometrical identifier in the tracking EDM, information about the geometry can be shared between the measurements and track parameters, and between the CPU and GPU.

The Kalman filter, detector geometry and magnetic field are global data shared by different tracks during the track reconstruction, and are therefore stored in the kernel global memory. In addition, four sets of track-specific data are required to execute the track fitting kernel on the GPU:
\begin{enumerate}
    \item Input measurements of the trajectory
    \item Starting track parameters to steer the track parameters propagation
    \item Track fitting configurations, e.g. a target surface to extract the fitted track parameters
    \item Fitted results including the fitting status, a collection of track states on the trajectory and the fitted track parameters on the target surface
\end{enumerate}
Track-specific data are allocated on the pinned memory on the host, i.e. page-locked memory, and this memory is allocated contiguously for each track. The number of detector surfaces intersected by the particle varies with the kinematics of the particle and the detector layout, i.e. the data loads are different between tracks. Managing this load imbalance would require dedicated memory management and task scheduling strategies. In this paper, the detector surfaces are therefore constructed to be boundless\footnote{Boundless surfaces always have an intersection with a track as long as the track is not parallel to the surface.} to guarantee that the same number of surfaces are traversed by the tracks. 
Each surface object requires approximately 120 bytes of memory. In addition to that, the size of memory allocated and transferred for different track-specific data for a detector with 10 plane surfaces is summarized in Table~\ref{Tab:data_size}. Considering the relatively large size of the track-specific data, and the fact that they are accessed only once during the track fitting, those data are also stored in the kernel global memory.
\begin{table}[h!]
\centering
\begin{tabular}{|p{0.6\linewidth}|r|} 
  \hline
   \textbf{Data type} & \textbf{Size (B)} \\
    \hline
    \hline
    Input measurements & 280 \\
    \hline
    Seeding parameters & 168 \\
    \hline
    Fitting configurations & 144 \\
    \hline
    Fitting status & 1 \\
    \hline
    Fitted states & 8480 \\
    \hline
    Fitted track parameters & 216 \\
    \hline
    \hline
    Total & 9289 \\
    \hline
\end{tabular}
\caption{The size (in bytes) of track-specific memory for a single track}
\label{Tab:data_size}
\end{table}

%% file: performance.tex
\section{Performance Evaluation}
\label{sec:perf}

The software performance is studied using a simple telescope-like detector geometry with 10 planar surfaces perpendicular to the global $x$ axis and placed equidistantly with 30 $mm$ between two adjacent planes. Realistic HEP track reconstruction applications typically involve a more complicated detector geometry.
A constant magnetic field of 2\,T along the global $z$ axis is used.
Samples containing a single muon per event are used for the performance evaluation. Muons are used for this study because they interact only minimally with the detector material and thus high quality track fits are expected. The muons have a transverse momentum uniformly distributed between 1 and 10\,GeV with both the azimuthal angle and polar angle fixed to zero.

\begin{figure}[!htb]
  \centering
  \includegraphics[width=1\linewidth]{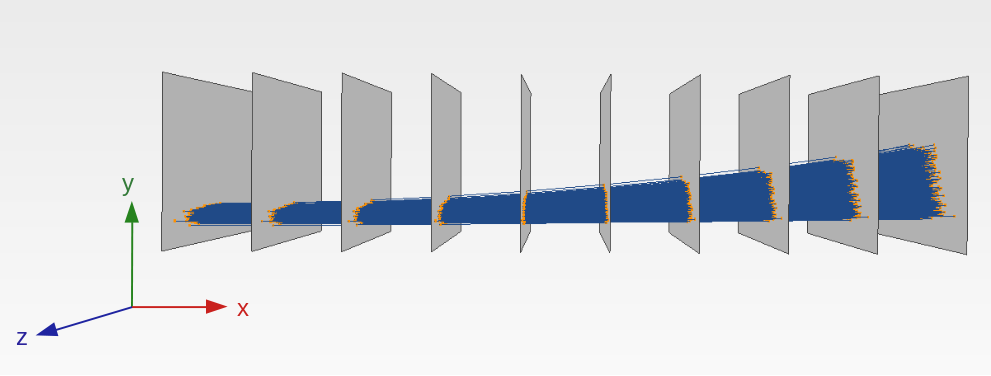}
  \caption{The detector configuration used to study the performance. It consists of 10 identical planar surfaces (gray planes) perpendicular to the $x$-axis. The trajectories of 10,000 simulated single muons (blue lines) and the associated simulated hits on the detector surfaces (orange dots) are indicated}
  \label{fig:vis_simulation}
\end{figure}

The Fatras fast simulation engine~\cite{Edmonds:1091969} within the ACTS toolkit is used to generate simulated hits of the single muons on the detector surfaces. Figure~\ref{fig:vis_simulation} illustrates the simulated hits on the detector surfaces for a sample of 10,000 single muons. The dense tracking environment at the HL-LHC is not expected to exceed 10,000 tracks per event.

As the pattern recognition algorithms required to find track candidates from measurements are beyond the scope of this paper, the known trajectories of the simulated particles are used for the fitting in place of track candidates provided by a pattern recognition step.
The measurements corresponding to the track candidates are obtained by smearing the positions of the simulated hits with Gaussian distributions to model detector resolution effects. A resolution of 50\,$\mu m$ in both the $x$ and $y$ dimensions of the detector, representative of the resolution of current pixel detectors at the LHC, is used. The initial set of track parameters for the track fit is based on the simulated particle vertex and momentum, smeared by Gaussian noise.

\subsection{Hardware and Software Environment}
Computing nodes of two supercomputers are used for running the performance tests:
\begin{enumerate}
    \item Cori Intel Xeon Haswell node, Cori Intel Xeon Phi Knight's Landing (KNL) node, and Cori GPU node at the National Energy Research Scientific Computing Center (NERSC)~\cite{cori-specs}
    \item Intel Xeon Skylake (SL) node and ATLAS-GPU01 node at the National Analysis Facility (NAF) at DESY
\end{enumerate}
Only GPUs from the Nvidia Tesla family are studied.
All nodes use the CentOS 7 operating system and all GPUs are using CUDA version 10.2.89.
Tables~\ref{Tab:cpu-config} and~\ref{Tab:gpu-config} show the detailed hardware and software configurations of the systems for CPUs and GPUs respectively.

\begin{table}[h!]
\begin{tabular}{|p{0.06\linewidth}|p{0.24\linewidth}| p{0.15\linewidth}|p{0.14\linewidth}|p{0.1\linewidth}|} 
  \hline
   \textbf{Sys.} & 
   \textbf{Model Name} & 
   \textbf{S$\times$C$\times$T} &
   \textbf{Clock Rate (GHz)} & 
   \textbf{Mem. (GB)} \\
    \hline
    \hline
    1 & Intel Xeon E5-2698 v3 \newline (Cori-Haswell) & 2x16x2 & 2.30 & 128 \\ 
    \hline
    1 & Intel Xeon Phi 7250 (Cori-KNL) & 1x68x4 & 1.40 & 96 \\
    \hline
    2 & Intel Xeon Gold 5115 (NAF-SL) & 2x10x2 & 2.40 & 376 \\
    \hline
    \end{tabular}
\caption{CPU configurations.
The Sys. column specifies whether the CPUs are used in the NERSC (1) or NAF (2) system.
The S$\times$C$\times$T column represents the number of sockets (S), cores per socket (C) and threads per core (T)}
\label{Tab:cpu-config}
\end{table}

\begin{table}[h!]
\begin{tabular}{|p{0.06\linewidth}|p{0.21\linewidth}|p{0.1\linewidth}|p{0.1\linewidth}| p{0.1\linewidth}| p{0.1\linewidth}|p{0.1\linewidth}|} 
  \hline
   \textbf{Sys.} & 
   \textbf{GPU} & 
   \textbf{FP32 Cores} & 
   \textbf{FP64 Cores} &
   \textbf{Clock Rate (GHz)} & 
   \textbf{Mem. (GB)}\\
    \hline
    \hline
    1 & GV100-SXM2 \newline (Cori-V100)& 5120 & 2560 & 1.53 & 16 \\
    \hline
    2 & GP100-PCIe \newline (NAF-P100)& 3584 & 1792 & 1.48 & 16\\
    \hline
    2 & GV100-SXM2 \newline (NAF-V100) & 5120 & 2560 & 1.53 & 32 \\
    \hline
\end{tabular}
\caption{GPU configurations.
The Sys. column specifies whether the GPUs are used in the NERSC (1) or NAF (2) system.
FP32 and FP64 columns denote the numbers of floating point compute units for single and double precision arithmetic operations, respectively}
\label{Tab:gpu-config}
\end{table}

These machines cover a wide range of architectures, with Cori-Haswell representing a standard compute node with two processors and a moderate amount of cores (for a total of 64 threads), while Cori-KNL features a higher core count due to the Xeon Phi's particular architecture (for a total of 272 threads).
Moreover, NAF-P100 and NAF-V100 allow two different GPU generations to be compared, with NAF-P100 being from the older Pascal architecture and NAF-V100 belonging to the newer Volta architecture.
The NAF-P100 is connected through a Peripheral Component Interconnect Express (PCIe) serial connector while the NAF-V100 uses the SXM2 connector, a multi-line serial connector that provides both Nvidia NVLink and PCIe connectivity~\cite{nvidia-doc}.
Cori-V100 is identical to NAF-V100 except for the lower amount of main memory.
Each of the GPUs contains multiple Streaming Multiprocessors (SMs), which are similar to CPUs.
However, each GPU is equipped with a large number of SMs, specifically, up to 60 SMs for the P100 and up to 84 SMs for the V100.
Each SM contains many CUDA Cores, which execute compute kernels in the form of threads (64 single precision/32 double precision cores per SM for both P100 and V100).
Each warp consists of 32 threads, with each warp running on one SM and each SM being able to execute up to 64 warps simultaneously.
Due to a large number of total cores, it is important to distribute work across a sufficient number of warps.
This allows the dedicated warp schedulers to achieve maximum utilization by keeping the cores busy with instructions.
While on P100 all threads share a single program counter (and therefore have to execute the same instruction at the same time), V100 manages execution state per-thread, allowing more independence.

\subsection{Tracking Performance}
\label{subsec:trk_perf}

The resolution of the Kalman filter-based track fit is validated by calculating the pulls of the track parameters as follows:
\begin{equation}
v_{pull} = \frac{v_{fit} - v_{truth}}{\sigma_{v}},
\end{equation}
where $v_{fit}$ and $\sigma_{v}$ are the value and uncertainty of the fitted track parameter respectively, and $v_{truth}$ is the corresponding parameter for the simulated particle.

\begin{figure*}[!htb]
  \centering
   \includegraphics[width=\textwidth]{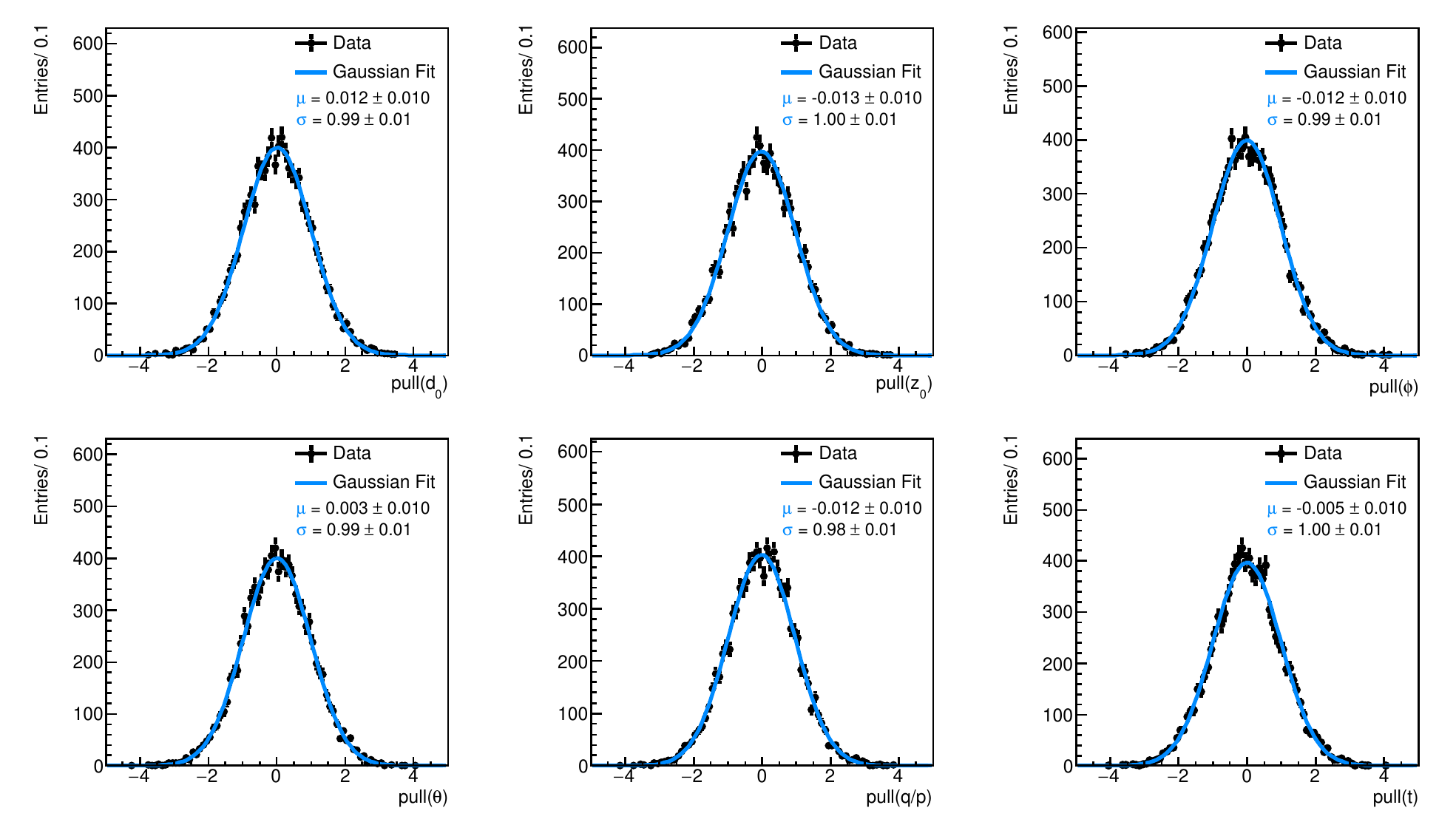}
  \caption{The distributions of the pull values of the fitted perigee track parameters, $(d_0, z_0, \phi, \theta, \frac{q}{p}, t)$, for a sample of 10,000 single muons obtained from Cori-V100. The black dots are the pull values, and the blue lines are Gaussian fits to the distributions}
  \label{fig:pull_muon}
\end{figure*}

The pull distributions of the six fitted perigee track parameters for a simulated sample of 10,000 single muons obtained from Cori-V100 are shown in Fig.~\ref{fig:pull_muon}. The pull distributions have means compatible with zero and widths compatible with one, which demonstrates that the track parameters and their uncertainties are estimated correctly by the track fit.

\subsection{Computing Performance}
\label{subsec:tech_perf}

The timing performance of the track fitting for different number of tracks on various computing architectures and configurations is measured. Ten tests are run for each measurement. The mean time of the ten tests is taken as the measurement result, and square root of the average of their squared deviations from the mean is taken as the measurement error.

The baseline tests are performed using only track-level parallelization, i.e.~without intra-track parallelization using shared memory. CUDA supports user configuration of the runtime properties of the GPU kernels, e.g. grid size and block size, and the launching of multiple kernels with multiple CUDA streams. Furthermore, the number of registers per thread can be controlled via the CUDA \verb!__launch_bounds__()! qualifier when the kernel is defined.
Unless explicitly specified, the tests are performed using one CUDA stream with 255 registers per thread as default configuration~\cite{nvidia-doc}.
Moreover, we choose a grid size of 5120$\times$1 to match the number of processing units on the V100 GPU, and a block size of 8$\times$8$\times$1, which is also the largest size of the matrix dealt with by the Kalman filter in ACTS.
Performance with intra-track parallelization and different CUDA configurations are also studied for comparison.

All the tests use single precision arithmetic operations with the exception of (a) the matrix inversion algorithm required by the smoother as detailed in Sect.~\ref{subsubsec:linear_algebra_support}, and (b) the explicit scenario which compares the different precisions described in Sect.~\ref{subsubsec:cpu_vs_gpu_section}.  

A singularity container with the executable and the required dependencies used to produce the results presented here is accessible in Ref.~\cite{xiaocong_ai_2021_4693389}.

\subsubsection{Performance of the Custom Matrix Inversion Algorithm}

\begin{figure}[!htb]
  \centering
  \includegraphics[width=1\linewidth]{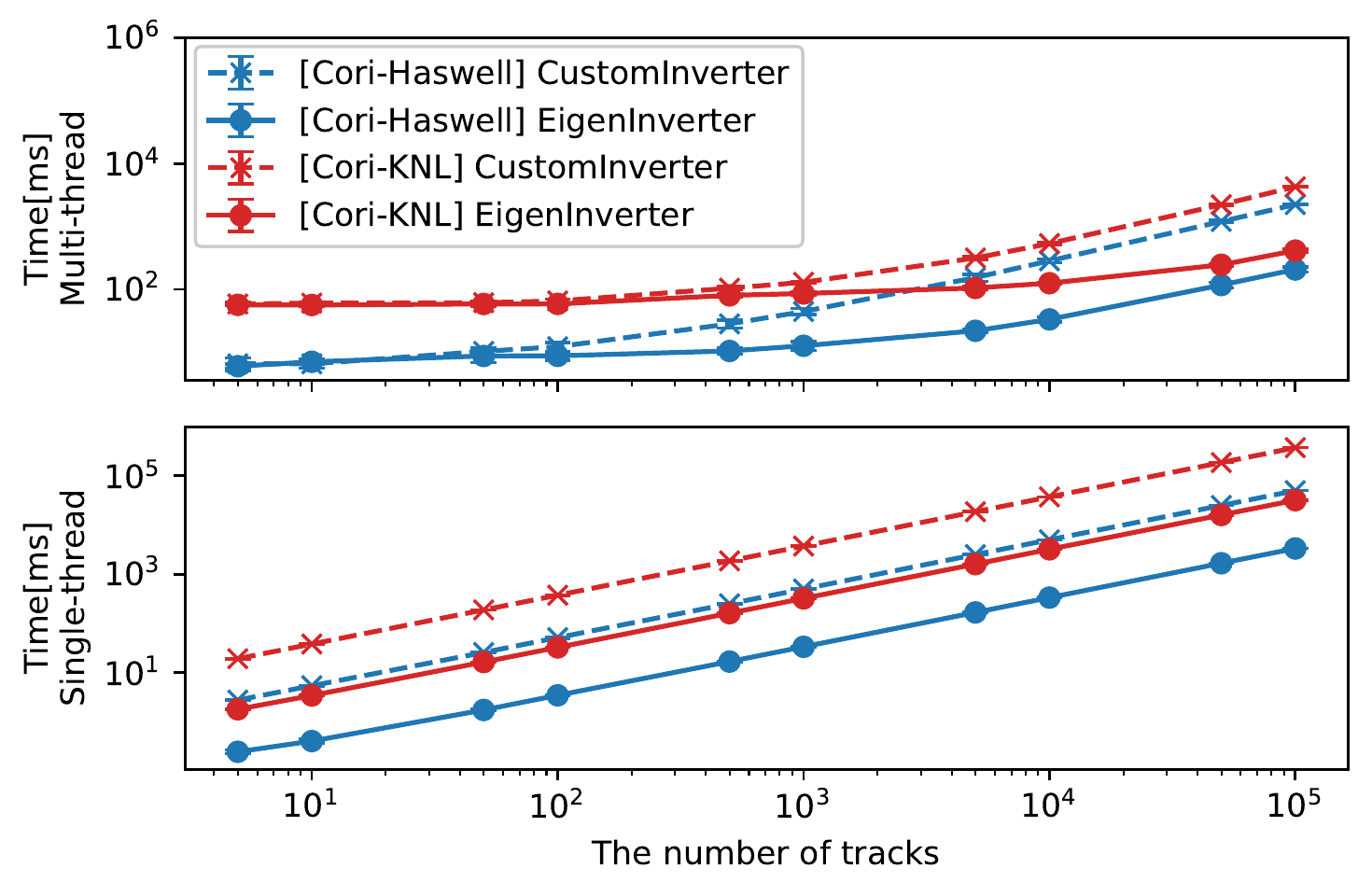}
  \caption{The fitting time as a function of the number of tracks with different matrix inversion algorithms on Cori-Haswell (dashed blue for custom matrix inversion, and solid blue for Eigen-based matrix inversion) and Cori-KNL (dashed red for custom matrix inversion, and solid red for Eigen-based matrix inversion). The top panel shows the results with 60 and 250 threads on Cori-Haswell and Cori-KNL, respectively, and the bottom panel shows the results with a single thread}
  \label{fig:inverters}
\end{figure}

Because the Eigen-based matrix inversion algorithm used by ACTS cannot be called inside CUDA kernels, a custom algorithm for matrix inversion implemented for this purpose is used.
Measurements are performed to compare our custom implementation to the Eigen-based implementation on the CPUs of both systems.
As shown in Fig.~\ref{fig:inverters}, it adds additional time to the fitting when the number of tracks exceeds 100.
While the Eigen-based implementation is significantly faster by a constant factor when using only one thread, this effect is much less pronounced when using as many threads.
Improving the performance of the custom matrix inversion operations on GPUs to match specialized linear algebra libraries would be expected to further improve the performance.

\subsubsection{Performance of CUDA Code on CPU}

\begin{figure}[!htb]
  \centering
  \includegraphics[width=1\linewidth]{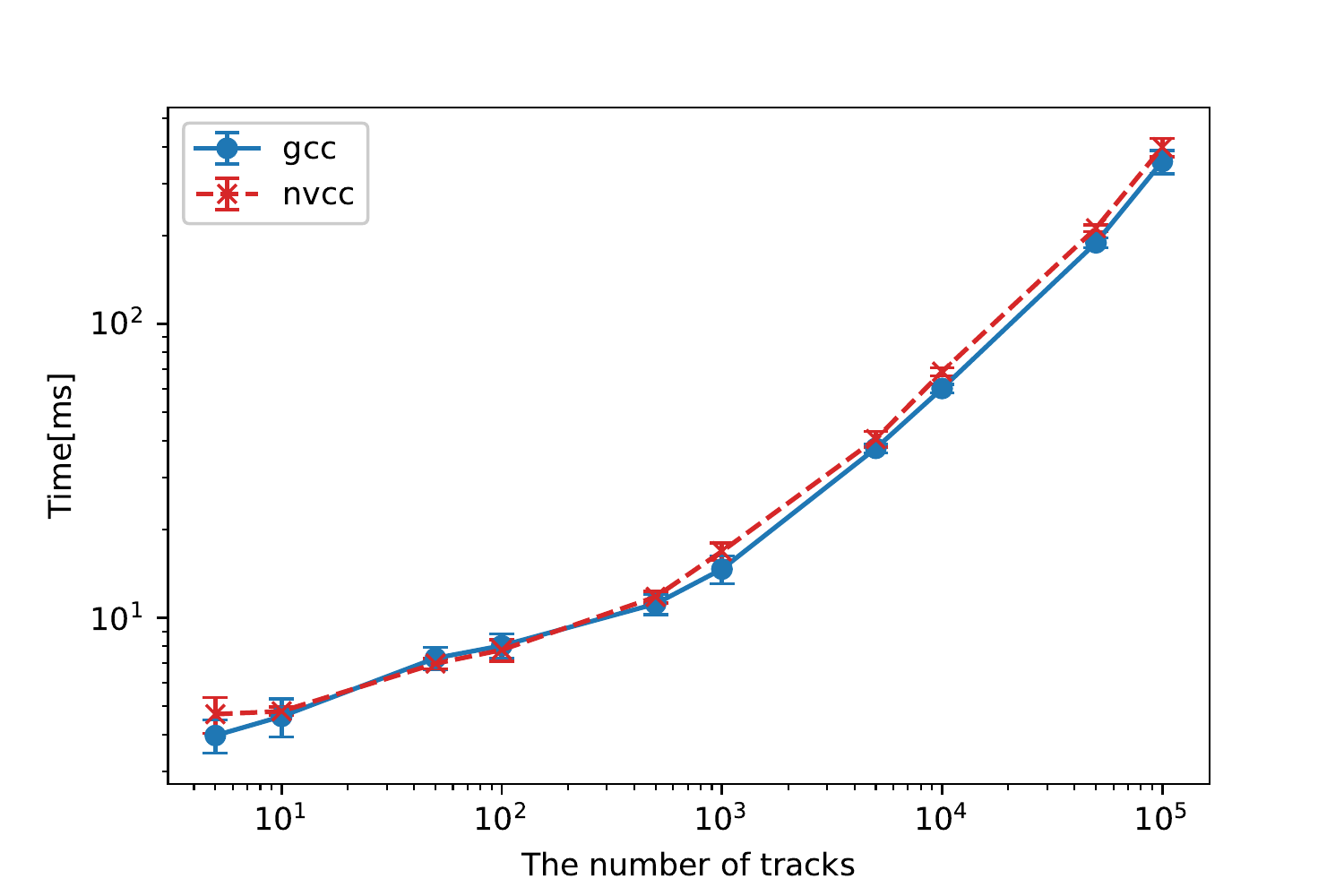}
  \caption{The fitting time as a function of the number of tracks on NAF-SL using 60 threads with Eigen-based matrix inversion obtained with the \texttt{gcc} (solid blue) and \texttt{nvcc} (dashed red) compiled executables, respectively}
  \label{fig:cuda_on_cpu}
\end{figure}

The GPU-based track fitting program compiled with \texttt{nvcc} can also run on CPUs (although the CUDA driver and CUDA runtime need to be accessible in order to successfully allocate page-locked memory on the host).
In this case, the track-level parallelization is achieved by using OpenMP threads and the host-device model including memory and execution offloading is bypassed.
The \texttt{nvcc} compiler uses the host compiler (\texttt{gcc} in this case) to generate the executable.
This approach ensures comparable execution time when the number of tracks is small but it induces a small performance penalty (between 4\,\% and 18\,\%) compared with the standard C++ implementation compiled with OpenMP support when the number of tracks exceeds 1000, as shown in Fig.~\ref{fig:cuda_on_cpu}.

Nevertheless, it demonstrates the potential for single-source code targeting heterogeneous hardware resources.
This is especially important for large and long-running software projects that might be used on different hardware architectures.
However, this still requires the code to be written using CUDA and implies the portability limitations discussed in Section~\ref{cuda}.

\subsubsection{Performance Comparison Between GPU Architectures}

\begin{figure}[!htb]
  \centering
  \includegraphics[width=1\linewidth]{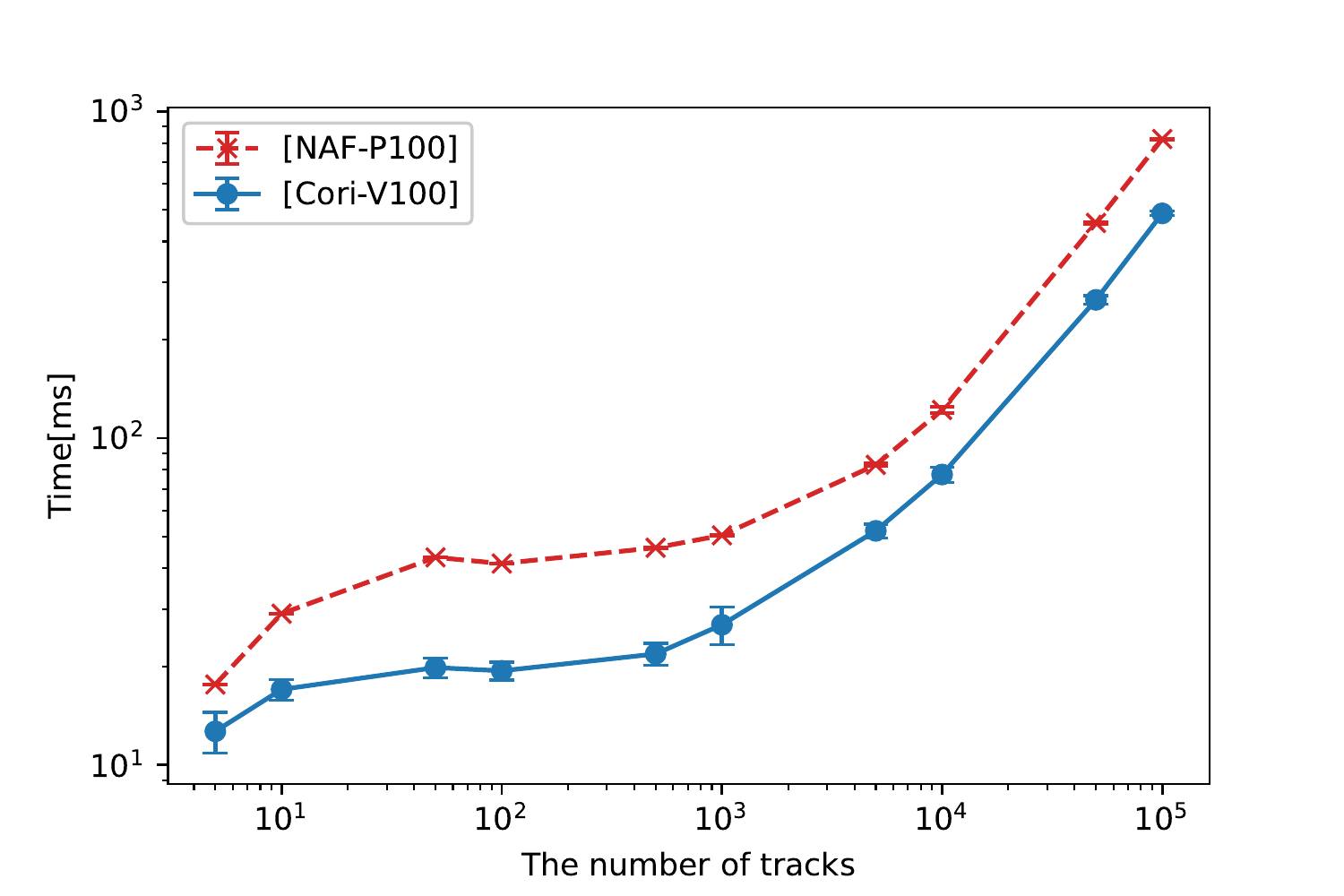}
  \caption{The fitting time as a function of the number of tracks on NAF-P100 (dashed red) and Cori-V100 (solid blue)}
  \label{fig:gpus}
\end{figure}

Performance of the Kalman filter-based track fitting on the P100 and V100 Nvidia Tesla cards, is compared.
Figure~\ref{fig:gpus} shows that fitting time on NAF-P100 is more than a factor of two longer than on Cori-V100, therefore the following tests will focus on the V100 due to its superior performance.
These performance differences are expected to a certain extent because the P100 and the V100 come from different hardware generations, with the V100 generally featuring more cores, higher clock rates and an improved interconnect.

\subsubsection{Performance Comparison Between CPU and GPU}
\label{subsubsec:cpu_vs_gpu_section}

\begin{figure}[!htb]
  \centering
  \includegraphics[width=1\linewidth]{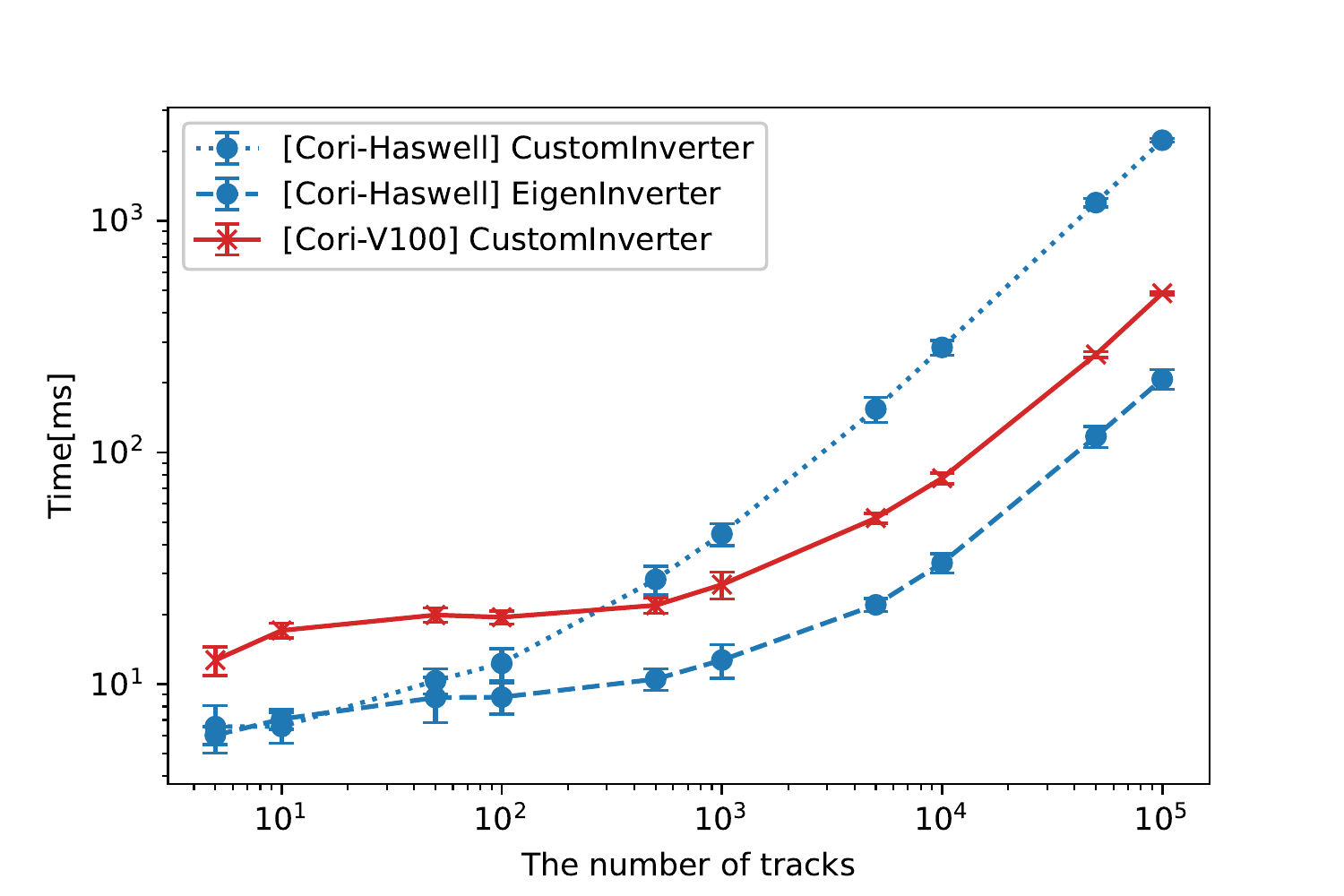}
  \caption{The fitting time as a function of the number of tracks on Cori-Haswell using 60 threads with Eigen-based matrix inversion (dashed blue) and custom matrix inversion (dotted blue), and on Cori-V100 (solid red)}
  \label{fig:cpu-vs-gpu}
\end{figure}

Figure~\ref{fig:cpu-vs-gpu} compares the fitting time on Cori-V100 with that on Cori-Haswell using different approaches to the matrix inversion.
As mentioned previously, only the custom matrix inversion is possible with Cori-V100.
When considering the custom matrix inverter, Cori-V100 displays superior performance compared to Cori-Haswell when the number of tracks exceeds 1000.
However, the Eigen-based implementation on CPUs still outperforms our custom inverter on GPUs, demonstrating that it is important to not only consider the potential benefits from porting the actual code to GPUs but to also take supporting libraries into account.

\begin{figure}[!htb]
  \centering
  \includegraphics[width=1\linewidth]{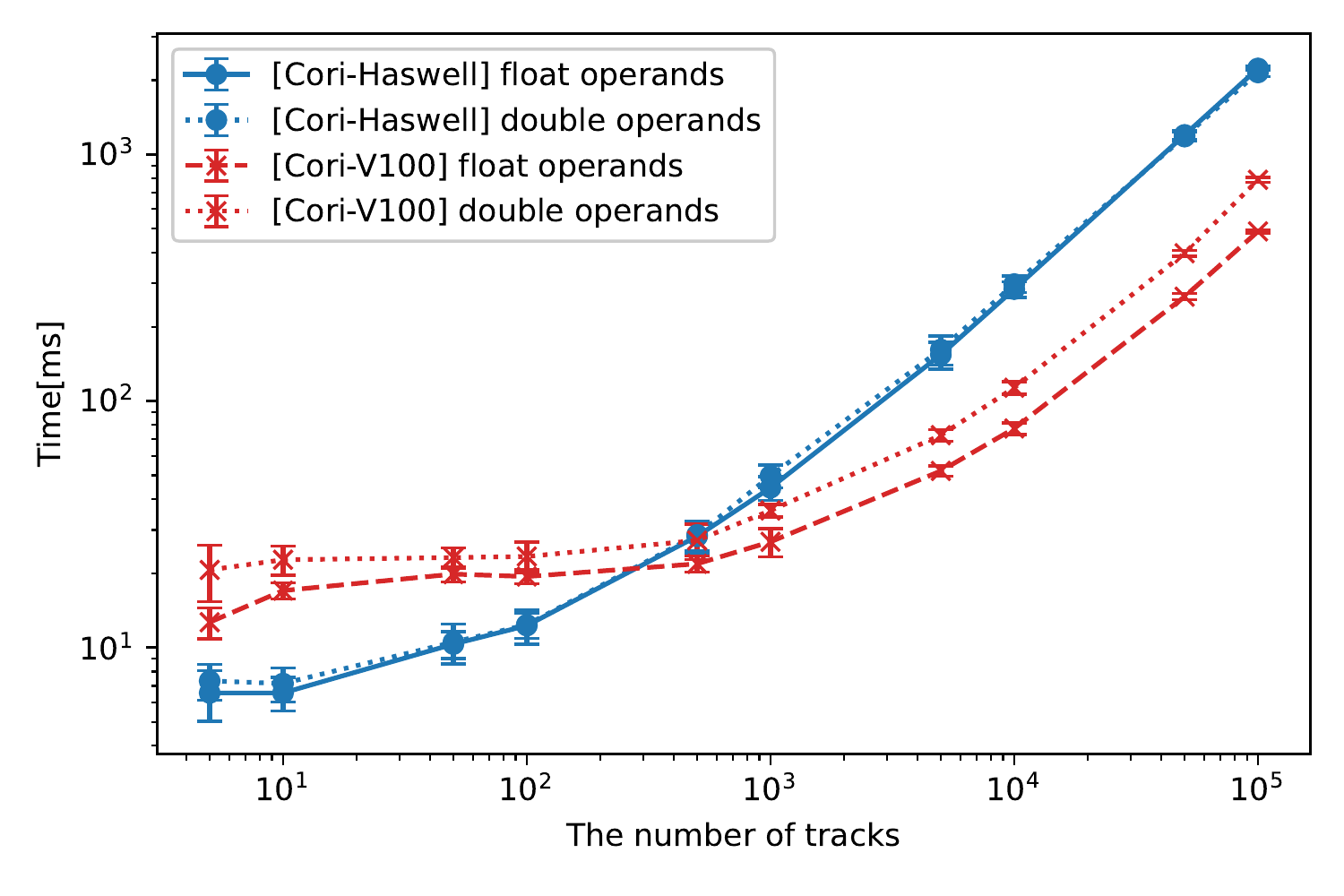}
  \caption{The fitting time as a function of the number of tracks using \texttt{float} and \texttt{double} operands on Cori-Haswell (solid blue for float, and dotted blue for double) and Cori-V100 (dashed red for float, and dotted blue for double). The tests are performed using the custom matrix inversion and the ones executed on Cori-Haswell are using 60 OpenMP threads}
  \label{fig:float_vs_double}
\end{figure}

Our matrix inversion algorithm performs all the operations in double precision regardless of the operands' types.
Figure~\ref{fig:float_vs_double} shows that there is little variation in performance between the different operands' types: when running the code on Cori-Haswell, there is virtually no performance difference for more than 1000 tracks, while double operands are slightly slower on Cori-V100.

As a consequence of parallel execution, the fitting time per track varies inversely with the number of tracks for all the considered platforms. In particular, for an HL-LHC scenario with up to 10,000 tracks, both the current prototype on GPUs and the Eigen-based implementation on CPUs show a fitting time per track in the order of microseconds.

\subsubsection{Performance with Different GPU Configurations}

The impact of usage of intra-track parallelization with shared memory, and variations in number of streams per device, grid size and block size as well as the number of fitted tracks are investigated and the most important results are discussed next. 

\begin{figure}[!htb]
  \centering
  \includegraphics[width=1\linewidth]{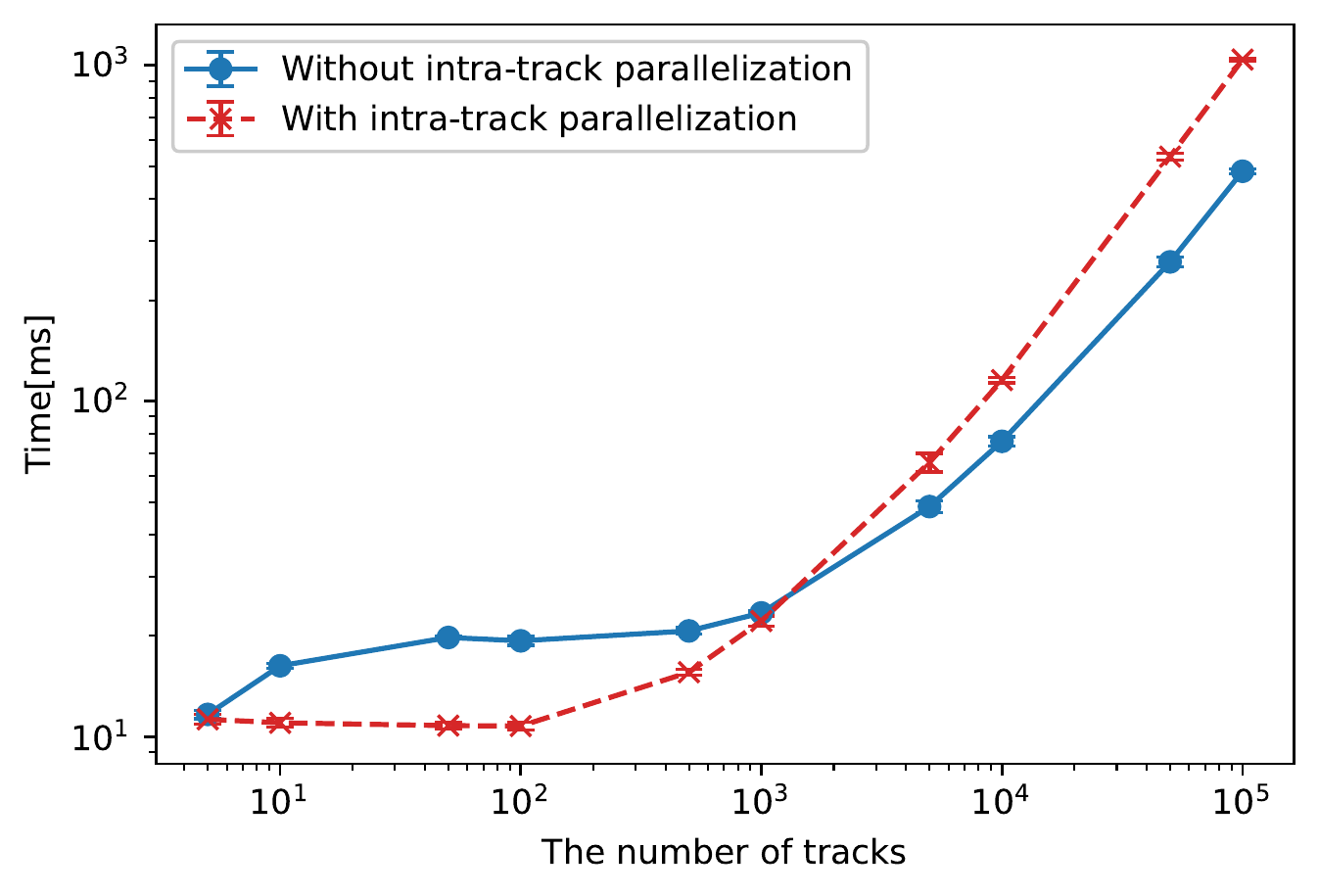}
  \caption{The fitting time as a function of the number of tracks with linear grid of size 100,000$\times$1 with (dashed red) or without (solid blue) intra-track parallelization on Cori-V100}
  \label{fig:memory-var}
\end{figure}

Figure~\ref{fig:memory-var} shows that performance gain could be achieved by using intra-track parallelization with shared memory when the number of tracks is below 1000. However, when the number of tracks exceeds 1000, using intra-track parallelization results in a performance penalty due to limitation of the available shared memory and resident threads per SM.

\begin{figure}[!htb]
  \centering
  \includegraphics[width=1\linewidth]{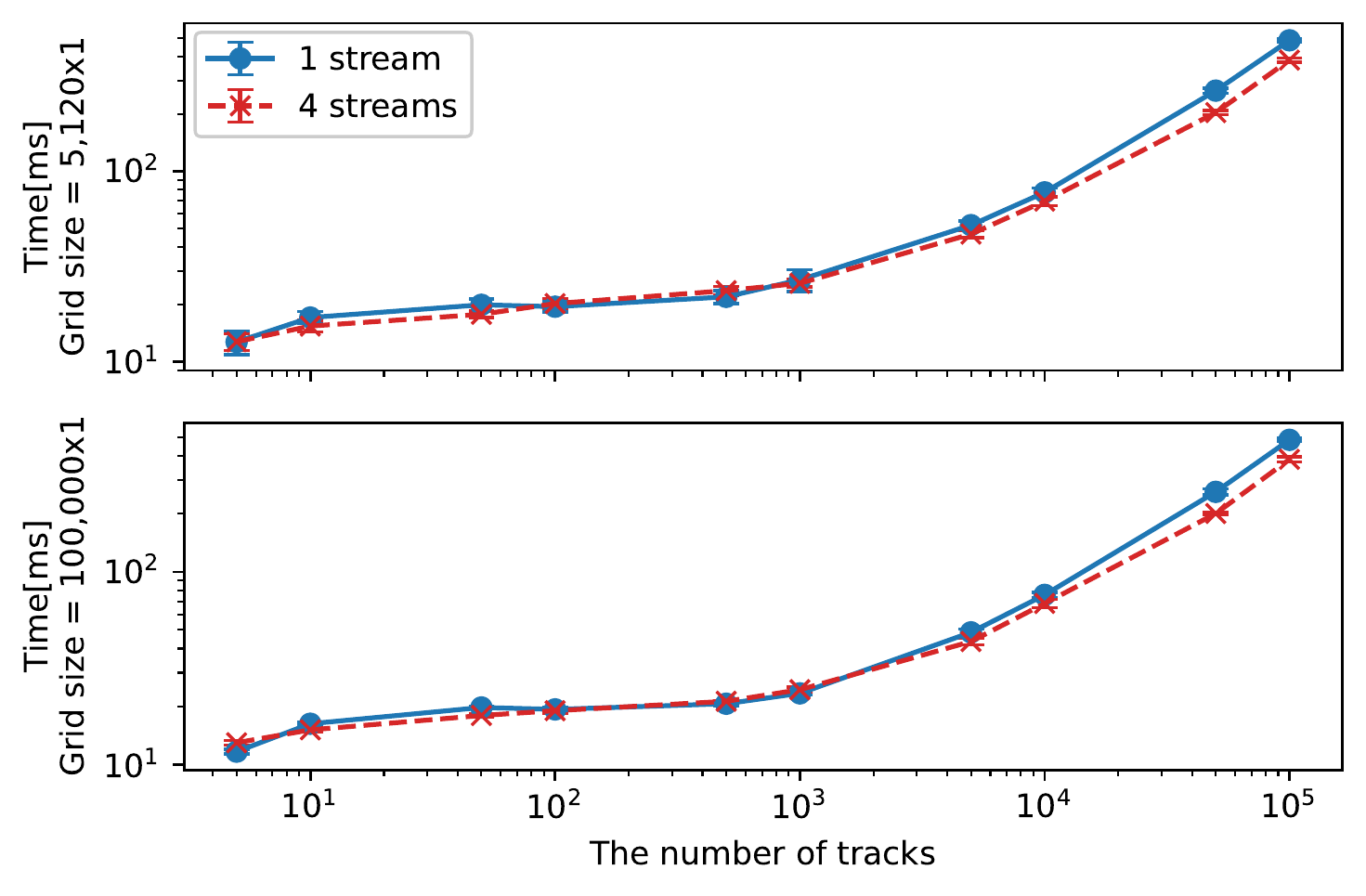}
  \caption{The fitting time as a function of the number of tracks with linear grids of sizes 5120$\times$1 (top) and 100,000$\times$1 (bottom), with one (solid blue) or four (dashed red) streams per device on Cori-V100}
  \label{fig:nStreams-var-float}
\end{figure}

Figure~\ref{fig:nStreams-var-float} shows that no significant performance gain is obtained from using more streams per device.  

\begin{figure}[!htb]
  \centering
  \includegraphics[width=1\linewidth]{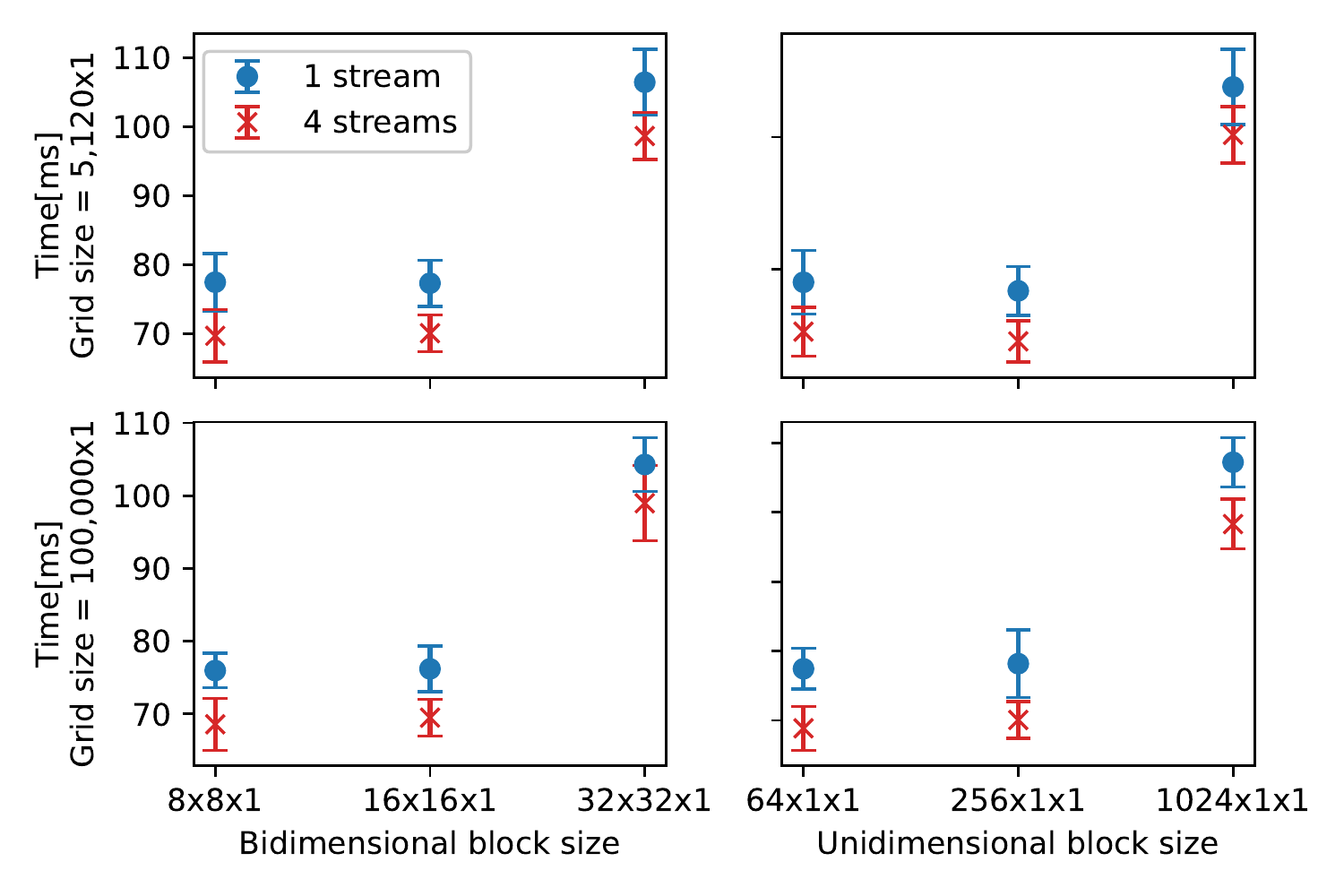}
  \caption{The fitting time for 10,000 tracks as a function of bidimensional (left) or unidimensional (right) block sizes with linear grids of sizes 5120$\times$1 (top) and 100,000$\times$1 (bottom), with one (blue) or four (red) streams per device on Cori-V100}
  \label{fig:grid-block-vars}
\end{figure}

Figure~\ref{fig:grid-block-vars} shows the required wall clock time with different grid and block sizes for 10,000 tracks. Note that when there are 1024 threads per block, the CUDA \verb!__launch_bounds__()! must be specified with the maximum number of threads per block no less than 1024 to avoid "too many resources requested for launch" errors. This results in a maximum of 64 registers per thread at Cori-V100 with 65,536 32-bit registers per SM.
While the performance differences between the two grid sizes are negligible, larger block sizes increase the runtime by up to a factor of 1.5.
These results show that it is important to choose block sizes that are appropriate for the underlying hardware, i.e.~when the overall workload is not large enough to saturate all the SMs on the GPU, a relatively large block size could lead to further imbalance of the workload distributed to the SMs and hence compromise the performance.

\begin{figure}[!htb]
  \centering
  \includegraphics[width=1\linewidth]{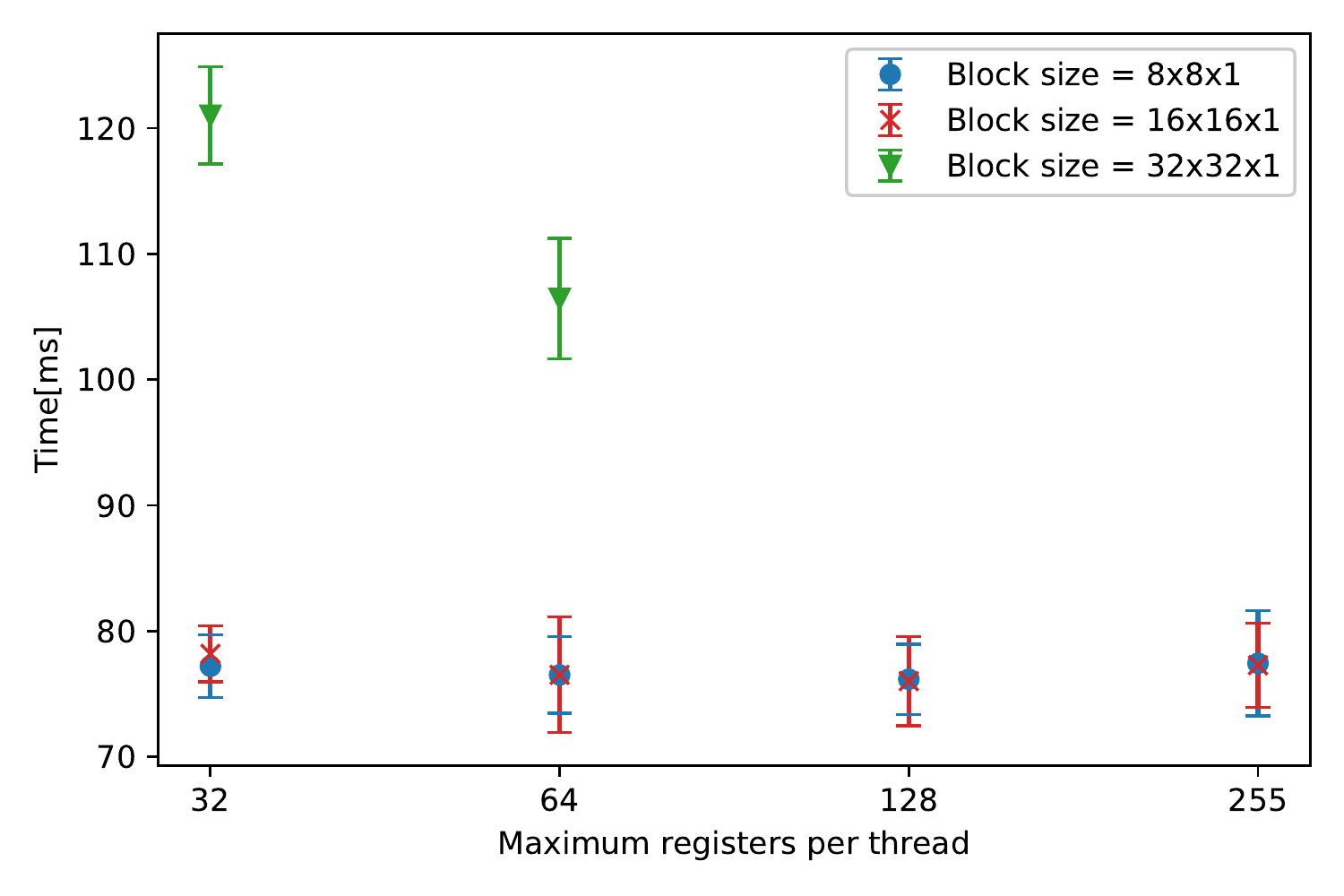}
  \caption{The fitting time for 10,000 tracks using various number of registers per thread and block sizes on Cori-V100 node. The circle blue, cross red and triangle green represent block sizes of 8$\times$8$\times$1, 16$\times$16$\times$1 and 32$\times$32$\times$1, respectively. When there are 1024 threads per block, a maximum of 64 registers per thread is allowed}
  \label{fig:max-registers}
\end{figure}

The number of registers per thread can be reduced with the goal of running more threads per block without exceeding the hardware limits (i.e. the maximum number of blocks per SM). Figure~\ref{fig:max-registers} shows how a variation in the number of registers per thread affects the overall performance for a track fitting workload of 10,000 tracks.
The performance varies little with the number of registers per thread. This is because the number of resident threads on the SMs are not increased by reducing the number of registers per thread when the overall workload is small. Increasing the block size can enforce at least the same amount of threads as the block size resident on some SMs, but this could lead to unwanted performance compromise due to inefficient GPU utilization. See Section~\ref{sec:disc} for further discussion.

\begin{figure}[h]
  \centering
  \includegraphics[width=1\linewidth]{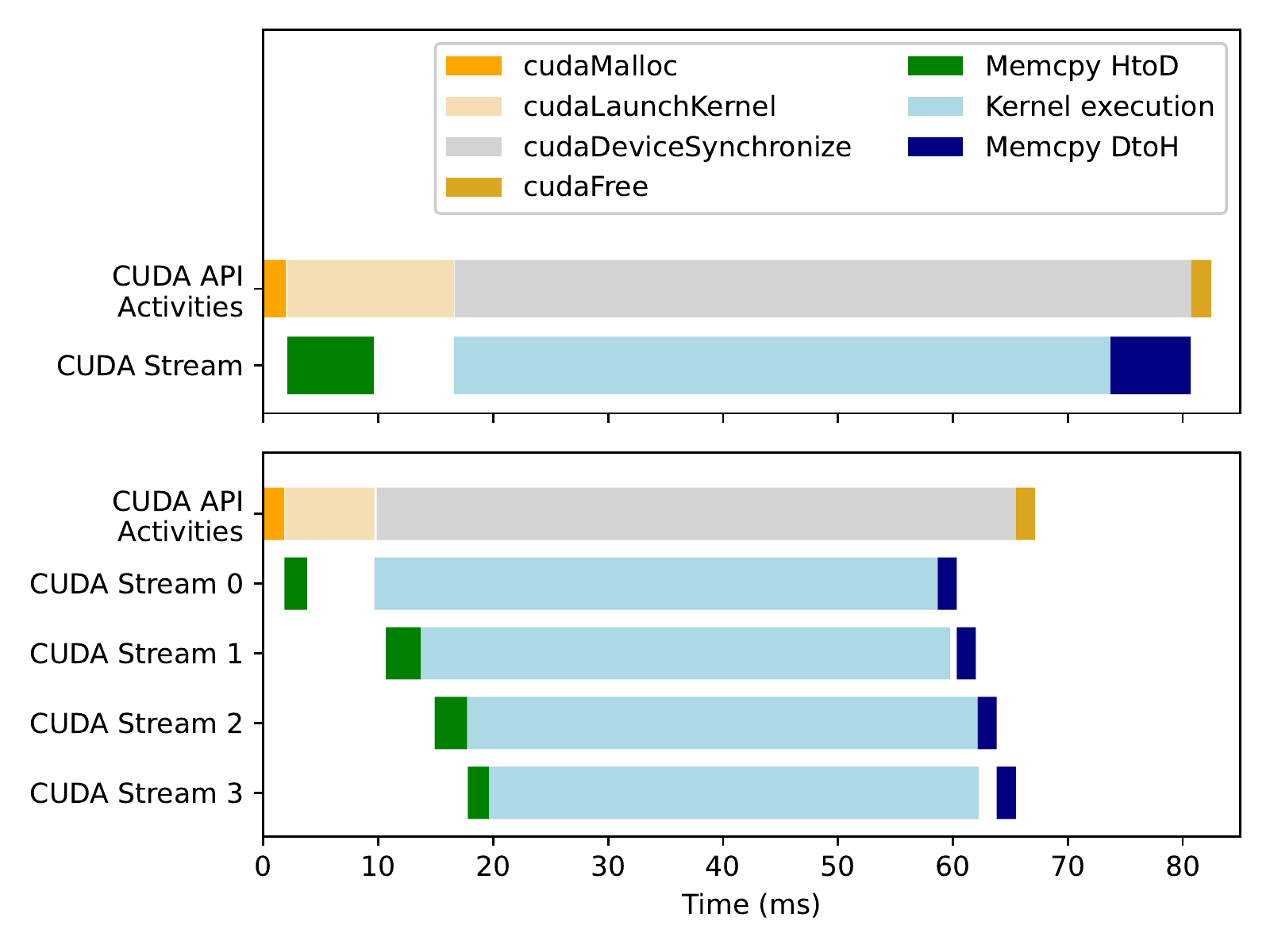}
  \caption{The timeline of CUDA activities for offloading track fitting for 10,000 tracks to Cori-V100 with one stream (top) and four streams (bottom). The starting times of memory allocation on GPU are taken as the $0$ point of the timelines}
  \label{fig:v100_timeline}
\end{figure}

The different components of the fitting time are analysed using Nsight Systems, one of Nvidia's new performance analysis tools~\cite{DBLP:journals/superfri/KnoblochM20}.
Figure~\ref{fig:v100_timeline} shows the timeline of those CUDA API activities required for GPU offloading, including the memory allocation on GPU, kernel launching, device synchronization and memory deallocation on the GPU, and the timeline of memory transfer and kernel execution in either one CUDA stream or multiple streams for 10,000 tracks.
When using only a single stream, kernel execution accounts for roughly 70\,\% of the total runtime, and memory transfer accounts for roughly 17\,\% with significant impact on the performance. The performance gain from overlapping the data transfer and kernel execution with multiple streams is limited by the memory transfer time. A dedicated CUDA synchronization method might improve this.

%% file: discussion.tex
\section{Discussion}
\label{sec:disc}
The timing performance studies in Sect.~\ref{sec:perf} use a single GPU. Potential gains in the timing performance by utilizing multiple GPUs and the GPU occupancy (which could have an impact on the timing performance) are investigated in addition.

\begin{figure}[!htb]
  \centering
  \includegraphics[width=1\linewidth]{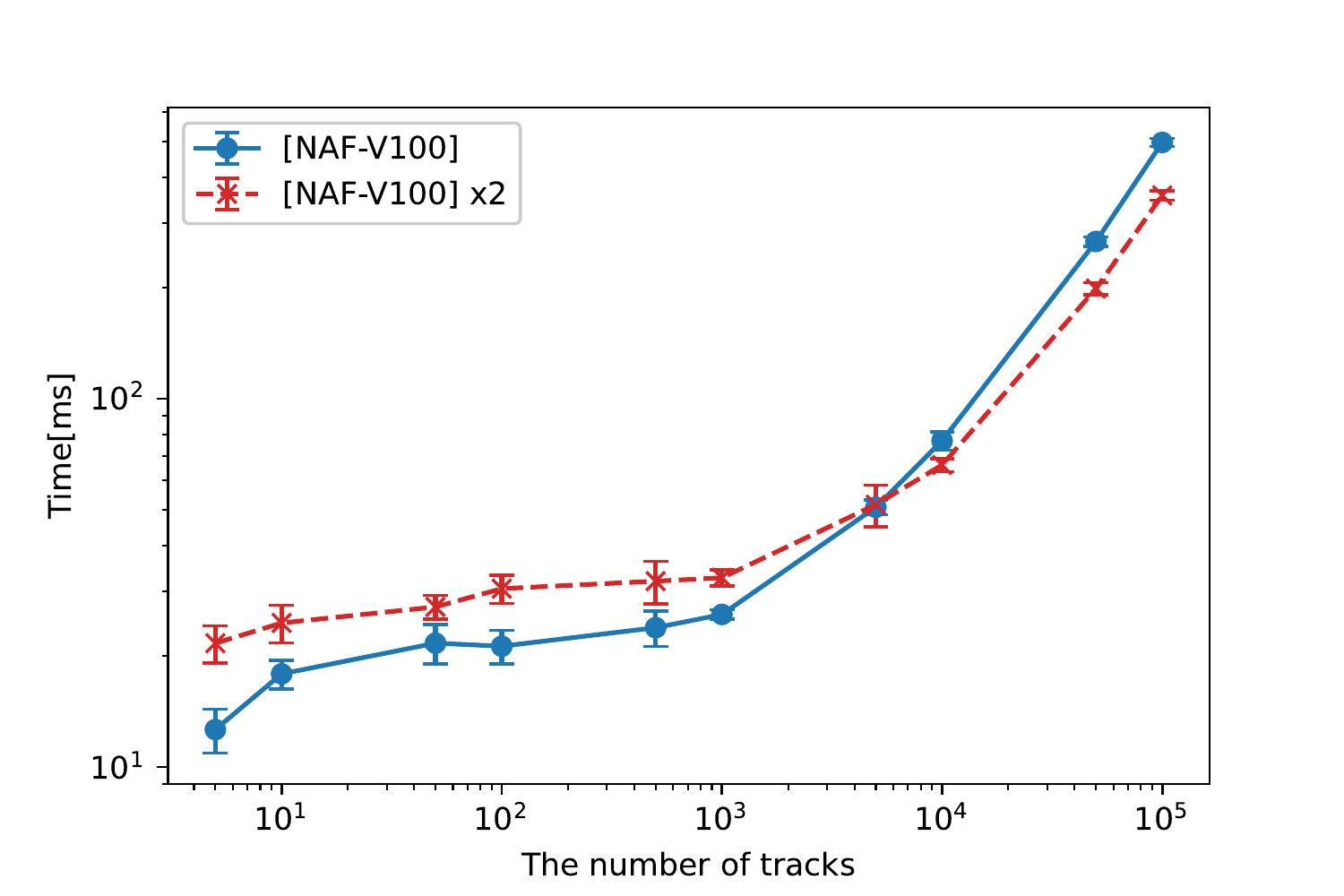}
  \caption{The fitting time as a function of the number of tracks executed in one stream per device with linear grid of size 5120$\times$1 and block size of 8$\times$8$\times$1, when using one NAF-V100 (solid blue) and two NAF-V100 in parallel (dashed red)}
  \label{fig:multi_gpu}
\end{figure}

While the performance of the GPU-based Kalman filter was tested on one GPU, the implementation can also run on multiple GPUs in parallel. The prototype uses a team of threads on the host, each one fitting the trajectory of a subset of tracks on a different GPU. Despite better problem size scaling, the multi-device solution has a slightly larger overall execution time for smaller numbers of tracks, as shown in Fig.~\ref{fig:multi_gpu}. Communication via Message Passing Interface (MPI) is required to fully exploit the parallelism by resolving synchronization overhead between the GPUs.

In addition, the latest versions of the Nvidia HPC Software Development Kit (SDK) provide new tools and libraries designed to maximize performance by optimizing memory transfers and scaling to multiple devices while targeting heterogeneous resources~\cite{nvidia-sdk}. Additionally, various studies regarding vendor-agnostic offloading approaches show promising results based on standard APIs and/or open-source, non-proprietary solutions~\cite{9309052,DBLP:conf/sc/GayatriYKD18}. These would be very interesting to explore in future iterations.

\begin{figure}[!htb]
  \centering
  \includegraphics[width=1\linewidth]{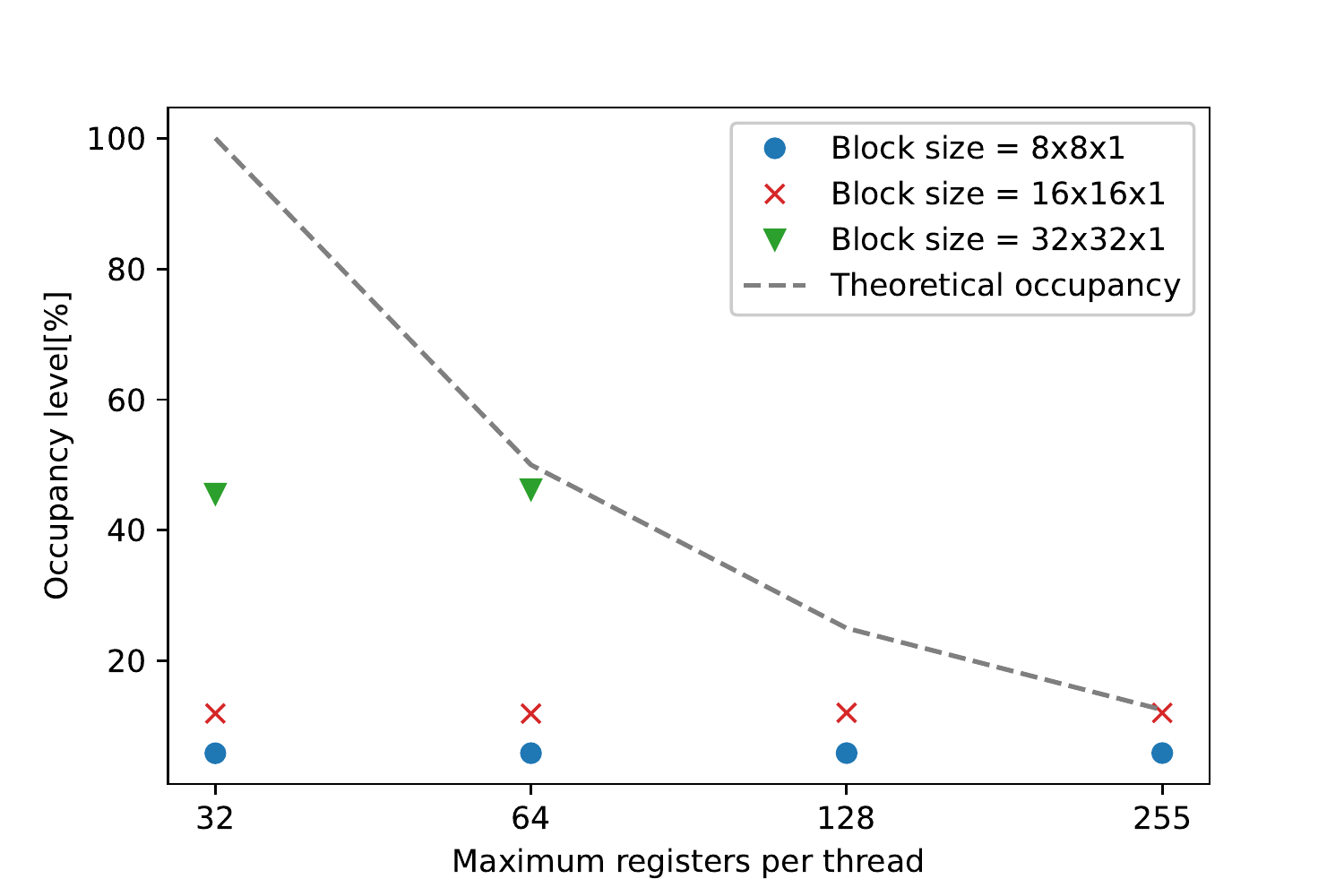}
  \caption{Warp occupancy levels using various number of registers per thread and block sizes for fitting 10,000 tracks executed in one stream with linear grid of size 5,120$\times$1 on Cori-V100. The black dashed line is the theoretical warp occupancy, and the circle blue, cross red and triangle green represent the achieved warp occupancy with block sizes of 8$\times$8$\times$1, 16$\times$16$\times$1 and 32$\times$32$\times$1, respectively. When there are 1024 threads per block, a maximum of 64 registers per thread is allowed}
  \label{fig:occupancy}
\end{figure}

The warp occupancy, defined as the ratio of active warps on an SM to the maximum number of active warps supported by the SM (e.g.~64 warps per SM on the V100), is analyzed using a different Nvidia performance  tool: Nsight Compute~\cite{DBLP:journals/superfri/KnoblochM20}.
Figure~\ref{fig:occupancy} shows the dependency of the theoretical warp occupancy and the achieved warp occupancy on the number of registers per thread and block size for a track fitting workload of 10,000 tracks. The theoretical occupancy is 100\,\% when the number of registers per thread is no larger than 32 and decreases when more registers are required for one thread, hence fewer threads are active. Since the maximum number of thread blocks per SM is 32 on the V100, the block size is not a limiting factor to theoretical occupancy as long as it is no less than 64. 

The achieved occupancy is well below the theoretical one, in particular when the block size is small. The reason is that an average of only 119 tracks are distributed to each SM of Cori-V100, providing a total workload of 10,000 tracks. Therefore, only 2 blocks are resident on the SM when the block size is 8$\times$8$\times$1 (128 threads), and at most 1 resident block when the block size is 16$\times$16$\times$1 (256 threads) or 32$\times$32$\times$1 (1024 threads). These correspond to an occupancy of 6.25\,\%, 12.5\,\%, and 50\,\%, respectively. In this case, reducing the number of registers per thread has no impact on the warp occupancy. Event-level parallelization is an effective approach to increase the workload, and hence improve the SM warp occupancy. This can be achieved by an offloading pattern with a fully contained chain of tracking modules that run on GPU requiring minimum data transfer between CPU and GPU~\cite{bocci2020heterogeneous}. The workload also needs to be accounted for when analyzing the impact of SM warp occupancy on the performance. For instance, better warp occupancy does not necessarily correspond to better timing performance, as shown in Fig.~\ref{fig:max-registers}, for the particular track fitting workload of 10,000 tracks studied here.
Further discrepancy between the achieved occupancy and the theoretical one arises from the imbalanced track fitting workload both within blocks and across blocks due to different momentum hence propagation paths between tracks. In this paper, the workload imbalances were already controlled by using a homogeneous detector geometry for all the tracks. For a realistic detector, the concept of tracking regions as presented in Ref.~\cite{Lantz_2020,Cerati_2020} can be used for the parallelization of track reconstruction. In the case of track finding, there is additional workload imbalance from the selection of compatible measurements from a pool of non-static number of measurements on a detector surface and possibly splitting the track propagation into multiple branches if there are more than one compatible measurement found. One possible approach to suppress the workload imbalance level would be to group the tracks based on their kinematic properties so that one group of tracks will encounter the same segmented detector region, and assign different groups of tracks to different grids or blocks. See Ref.~\cite{Lantz_2020,Cerati_2020} for further discussion.

%% file: conclusion.tex
\section{Conclusion}
\label{sec:conclusion}

The reconstruction of charged particle trajectories for current and future high-energy physics experiments is a significant computational challenge. New approaches are needed to cope with the dramatically increased event complexity and rates and with the movement away from x86 architectures. The Kalman filter algorithm is the mainstay of current track reconstruction strategies. 
We presented a proof-of-concept implementation of a full Kalman filter track fitting algorithm using ACTS on two different Nvidia GPU architectures using a simplified detector geometry and a constant magnetic field. We have performed studies of its physics and technical performance and compared this to results using CPUs with a particular focus on the limitations observed. Ideas for improvements for future implementations were discussed.

As existing fast matrix inversion algorithms cannot run on GPUs, we developed a custom prototype matrix inversion algorithm, which does not match the CPU performance of highly-optimized state-of-the-art algorithms. When controlling for the matrix inversion algorithm, worse performance for low track multiplicity is obtained with the GPUs compared to the CPUs and the performance is improved by a factor of up to 4.6 with respect to the CPUs for events with more than 1,000 tracks. Significant performance differences are shown between the different GPU architectures.

Parallelization within the track fit was implemented and performance gain was observed with a relatively low track multiplicity. The performance dependence on GPU configurations was also studied. The performance was largely independent of the grid size and did not change when using multiple kernels. Memory transfer and other overhead can account for up to 30\,\% of the total run time for the track fit.

The typical HL-LHC track multiplicity of 10,000 tracks per event is a relatively small workload for GPUs. For events with 10,000 tracks, performance gain by up to a factor of 1.5 is achieved by using a smaller block size. The small workload is also the main limiting factor in the achieved occupancy on the GPU.

 We have compared different methods for the Kalman filter implementation and studied the dependence on the GPU configuration. We have identified limitations of the approach and highlighted areas for future work directions.

Specifically, an evaluation of alternative approaches for GPU offloading, especially those provided by vendor-agnostic interfaces such as OpenMP, can be expected to result in improved performance portability.
Moreover, further improvements to the GPU-based matrix inversion algorithm can be expected to bring its performance closer to existing CPU implementations.